\documentclass[aps,pre,twocolumn,groupedaddress]{revtex4-1}
\usepackage{graphicx}
\usepackage{color}
\usepackage{amssymb}
\begin{document}

%\preprint{}

\title{Machine Learning Characterization of Structural Defects in Amorphous Packings of Dimers and Ellipses}


\author{Matt Harrington}
\email{mharrin@sas.upenn.edu}
\author{Andrea J. Liu}
%\email{ajliu@physics.upenn.edu}
\author{Douglas J. Durian}
\email{djdurian@physics.upenn.edu}
\affiliation{Department of Physics and Astronomy, University of Pennsylvania, Philadelphia, Pennsylvania 19104, USA}

%\thanks{}


\date{\today}

\begin{abstract} 
Structural defects within amorphous packings of symmetric particles can be characterized using a machine learning approach that incorporates structure functions of radial distances and angular arrangement. This yields a scalar field, \emph{softness}, that correlates with the probability that a particle is about to rearrange. However, when particle shapes are elongated, as in the case of dimers and ellipses, we find the standard structure functions produce imprecise softness measurements. Moreover, ellipses exhibit deformation profiles in stark contrast to circular particles. In order to account for effects of orientation and alignment, we introduce new structure functions to recover predictive performance of softness, as well as provide physical insight to local and extended dynamics. We study a model disordered solid, a bidisperse two-dimensional granular pillar, driven by uniaxial compression and composed entirely of monomers, dimers, or ellipses. We demonstrate how the computation of softness via support vector machine extends to dimers and ellipses with the introduction of new orientational structure functions. Then, we highlight the spatial extent of rearrangements and defects, as well as their cross-correlation, for each particle shape. Finally, we demonstrate how an additional machine learning algorithm, recursive feature elimination, provides an avenue to better understand how softness arises from particular structural aspects. We identify the most crucial structure functions in determining softness and discuss their physical implications.
\end{abstract}

%\keywords{}

\maketitle

\section{Introduction}
\label{sec:Intro}

Under sufficiently strong mechanical loads, particulate constituents of solid materials deform locally, exhibiting failure in the form of rearrangements. The locations of initial rearrangements are often dictated by the presence of structural defects, i.e. weakness, in the underlying structure. In crystalline solids, structural defects, such as dislocations, are immediately apparent with full knowledge of the atomic structure. Local plastic events at the sites of defects can lead to dislocation motion, producing bulk scale deformation~\cite{Taylor1934}. Disordered solids, such as metallic glasses, do not exhibit structural defects in the same way. While systems with amorphous structure often exhibit short-range order, the degree of regularity in structure decreases sharply at distances on the order of the size of a few constituents and larger. Still, disordered solids under stress can undergo localized rearrangements~\cite{Argon1979,ArgonBubbleRafts1979}, which can coalesce into system-spanning failure events, such as shear bands and fractures. Plastic rearrangements in disordered solids are also thought to originate from flow defects, often referred to as shear transformation zones~\cite{FalkLangerPRE1998,FalkLangerARCMP2011}. A broad spectrum of modern materials are classified as disordered solids, so better understanding their structural defects is crucially important to improving material design.

To better understand how disordered solids deform locally, many researchers have chosen to focus on disordered nanoparticle assemblies, colloidal packings, and granular materials. These systems are distinct from atomic systems, in part because one can directly image the underlying structure and how it changes under mechanical or thermal forces using microscopy or photography. Typical particle sizes in these systems range from $10^{-9}$ m to $10^{-2}$ m, far above the atomic scale of constiuents in metallic glasses. While this is a wide range of length scales with varying local interactions, there exist universal properties of disordered solids. For instance, a number of systems from the classes specified above, along with literature-curated results for metallic glasses, exhibit a common yield strain of $3\%$~\cite{CubukIvancicScience2017}. Other efforts explore the presence of avalanches, interspersed drops in stress, in systems spanning from the nano scale to the geological scale~\cite{AvalancheSciRep2015}. These suggest that insight into deformation behavior in disordered solids can be elucidated by model systems that offer readily available measurements of structure.

Granular materials represent a class of disordered solids in which the full structure can be observed experimentally using 2D geometries or novel 3D imaging techniques~\cite{GranularImagingReview2017}. They are ensembles of discrete macroscopic particles in which thermal fluctuations are negligible and interactions of dry grains are dominated by dissipative and repulsive contact forces~\cite{JaegerNagelBehringerRMP1996}. Prior work on identifying structurally weak regions has largely been done in the context of simulated systems, in which interactions between grains can be explicitly modeled. Specifically, low-frequency vibrational modes can identify marginally stable particles that undergo rearrangements~\cite{BritoJSM2007,ManningLiuPRL2011,SchoenholzPRX2014}. Recent studies of relaxation events in simulated glassy systems have built on this using similar approaches, including vibrational mean-square displacements~\cite{DingNatComms2016}, local thermal energy~\cite{ZylbergPNAS2017}, and local yield stress~\cite{PatinetPRL2016,BarbotPRE2018}. In the context of experimental granular systems, one must also contend with the presence of body friction, so a characterization of vibrational modes is generally not possible. In these contexts, researchers have pointed to structural anisotropies to draw correlations with dynamics. Indeed, free volume has long been identified as an indicator of structural defect and plasticity in glassy systems~\cite{Turnbull1961,Spaepen1977}. While free volume can be difficult to precisely measure in granular systems, quantities reflecting it readily arise from Voronoi tessellations~\cite{SlotterbackPRL2008,SchroderTurkEPL2010,SchroderTurkNJP2013,MorseCorwinPRL2014,RieserPRL2016}. However, it has been difficult to identify a singular component of structural anisotropy, absent knowledge of local interactions, that is highly predictive of dynamics.

To define a predictive structural signature, not only for granular materials but for a wide array of disordered systems, researchers have devised and implemented a machine learning technique that computes a single structural parameter that strongly correlates with rearrangement probability~\cite{CubukSchoenholzPRL2015,SchoenholzCubukNatPhys2016,CubukSchoenholzJPCB2016,SchoenholzCubukPNAS2017,SussmanSchoenholzPNAS2017,CubukIvancicScience2017,Sharp2018,Ivancic2018}. This parameter is known as \emph{softness} and arises from a support vector machine (SVM) that computes a hyperplane best separating rearranging and non-rearranging particles, given a multitude of structural features ascribed to each particle. While the SVM is often used in practice to model binary classifications, softness is determined by the signed distance from the hyperplane to the particle's location in structural feature space. Thus, softness acts a continuous quantity whose magnitude is expected to reflect the susceptibility of a particle to an imminent rearrangement.
%Something important about the choice of how we do a classification algorithm, but really care about the signed distance softness, is that we don't want this method to be purely a black box or oracle. Instead, we want the model to return something physical, which in this case is a distillation of local structural features, with regard to susceptibility to rearrangement. Maybe a CNN could do better prediction-wise, but you lose physical intuition that can be gained.
%{Important to note: Softness is not some `oracle' or `black box' that is intended to be 100\% predictive of rearrangements; instead, it is a a distillation of structural components that, taken as a whole, are susceptible to rearrangements. This study has illuminated strategies for how to extend this approach to constituents with anisotropic interactions, as well as identified a new avenue for physical interpretation which examines the subset of critical structural components that determine a particle's rearrangement susceptibility.}

So far, softness has not been computed for systems with anisotropic particle shapes. In natural and industrial granular materials, grains are rarely spherical, so a large sector of research is dedicated just to the structure and rheology of nonspherical grains~\cite{BorzsonyiSM2013}. Focusing on solid packings of nonspherical particles, there is ongoing work to characterize packing anisotropies using Voronoi-based techniques mentioned earlier~\cite{SchallerEPL2015}. Existing studies have also considered different structural aspects of dense packings of dimers, ellipse, and other elongated particles~\cite{MailmanPRL2009,SchreckSoftMatter2010,ShenPRE2012,SchreckPRE2012,MarschallPRE2018,VanderWerfPRE2018}. Shape dependence of the mechanical response and strength of granular solids is especially well-known and under ongoing study~\cite{UniaxialNonspherical2012,PrintingGrainsNatureMat2013,PrintingGrainsSoftMatter2014,UniaxialNonspherical2016,Xie2017,HarringtonPRE2018,Murphy2018}. Particle shape effects are not limited to just granular systems, however, and indeed extend into other classes of disordered solids. For example, elongated particles in disordered nanoparticle assemblies drasitically increase the fracture toughness of the system, suppressing the appearance of cracks and shear bands during standard indentation tests~\cite{ZhangACSNano2013}. It stands to reason, then, that softness should be explored for anisotropic particle shapes as well. This is an area we would like to address, specifically in the case of a 2D granular material representing a model disordered solid.

In this article, we begin by describing the granular system and the three particle shapes we use to explore symmetric and elongated shapes in Section~\ref{sec:Methods}. In Section~\ref{sec:SVM}, we define the established SVM approach to computing softness that has previously been applied to a variety of disordered solids. We characterize the performance of this algorithm on all three shapes, highlighting shortcomings in the case of elongated shapes. These deficiencies are remedied in Section~\ref{sec:AdjFeats}, in which we define a new set of structure functions that perform better with dimers and ellipses. In Sections~\ref{sec:SpatialCorr} and~\ref{sec:RFE}, we provide physical interpretations that we derive from softness and our overall machine learning approach. Specifically, Section~\ref{sec:SpatialCorr} examines the spatial extent of the rearrangement and softness fields, as well as the cross-correlation between them. Then, Section~\ref{sec:RFE} describes how another machine learning algorithm, recursive feature elimination, highlights structural aspects that are most---and least---critical in determining the softness of a particle. We conclude with a discussion of the results and descriptions of prospective studies in Section~\ref{sec:Discussion}.

\section{Materials and Methods}
\label{sec:Methods}
\begin{figure}
 	\includegraphics[width=\linewidth]{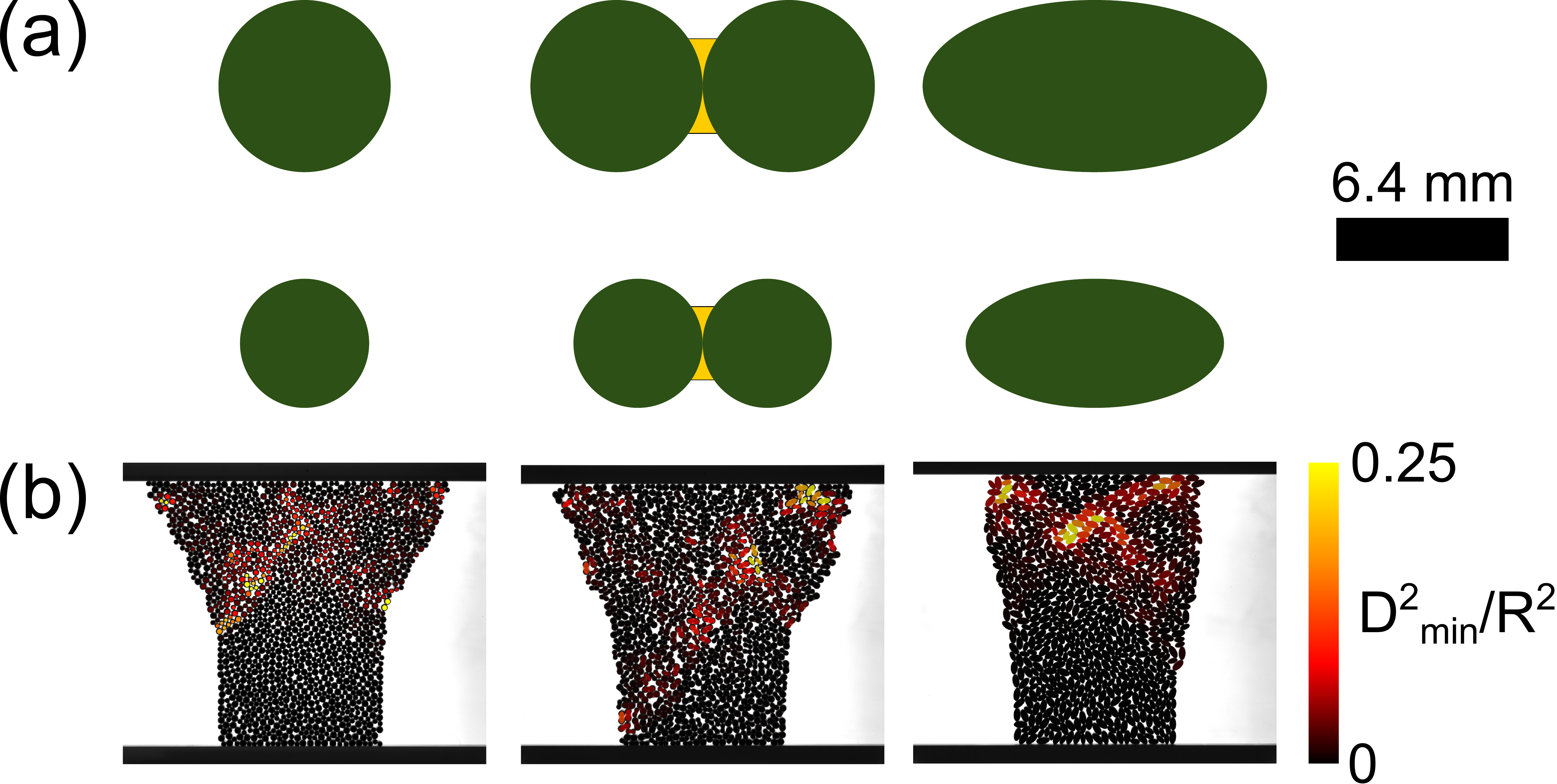}
	\caption{\label{fig:Particles_D2min} (a) Top-down sketches of the particles used in this study: (left) monomers, (center) dimers, with adhesive exaggerated for clarity, and (right) ellipses. The length of the scale bar (6.4 mm) corresponds to the large monomer diameter. (b) Raw images of the pillar with particles colored by their $D^2_{\mathrm{min}}$ value, according to the color bar. For monomers and dimers, the total compressive strain applied is $\gamma \sim 0.2$, while for ellipses, $\gamma \sim 0.1$. Full videos of a pillar compression trial with overlaid $D^2_{\mathrm{min}}$ are available in the Supplemental Material~\cite{Supplement}.} 
\end{figure}

Our experimental system is a two-dimensional granular pillar driven by uniaxial compression, using an apparatus previously described in Refs.~\cite{CubukSchoenholzPRL2015,RieserLangmuir2015,LiRieserPRE2015,RieserPRL2016,RieserThesis,HarringtonPRE2018}. The granular system consists of bidisperse rods that stand upright on a table-top acrylic substrate. The ratio of large particle number to small particle number is 1:1. Three particle shapes are used, monomers, dimers, and ellipses, as illustrated in Fig.~\ref{fig:Particles_D2min}(a). The monomers and dimers used in this study are the same as those described in Ref.~\cite{HarringtonPRE2018}. Monomers are acetal (Delrin) rods with large and small diameters of 0.25 in. (6.4 mm) and 0.1875 in. (4.8 mm), respectively, and a uniform height of 0.75 in. (19 mm). Dimers are pairs of like-sized rods bonded together with a cyanoacrylate adhesive. Ellipses are fabricated using 3D-printing (ProJet 3500 HD, 3D Systems) of a UV-cured acrylate material (VisiJet X). Large and small ellipses have the same lateral dimensions and aspect ratio as large and small dimers, respectively. While the height of all dimers is 0.75 in. (19 mm), all ellipses have a height of 0.375 in. (9.5 mm).

% <Update the figure captions as requested; Fig 1B – Monomer img = Run 1 of Training Set, Frame 5759 -> dy ~ 2.25 in. (\gamma ~ 0.2); Dimer img = Run 1 of Training Set, Frame 6519 -> dy ~ 2.60 in. (\gamma ~ 0.2); Ellipse img = Run 1 of Training Set, Frame 3439 -> dy ~ 1.34 in. (\gamma ~ 0.1)>

An ensemble of grains, all of one shape, are arranged into a tall, narrow pillar with an approximate aspect ratio of 2:1 using a rigid frame. The initial pillar width, $W_0$ = 6 in. (15.2 cm), is kept constant for all trails. The initial pillar height, $H_0 \sim$ 12 in. (30.4 cm), varies between trials and particle shapes. The total number of particles is determined by whether we are using symmetric particles (monomers) or elongated particles with aspect ratio 2 (dimers, ellipses). $N = 1500$ for monomers, while $N = 750$ for dimers and ellipses, in order to keep the pillar dimensions consistent. The pillar is unaxially compressed from the top by a slowly moving bar ($v_c$ = 0.0033 in./s = 85 $\mu$m/s), while a static bar is in contact with the pillar bottom. 
% The initial area fraction of ellipses is $0.866 \pm 0.005$; I don't have explicit measurements for these runs of monomers and dimers currently, but from the PRE 2018, monomers (N = 1000) have $0.823 \pm 0.004$; dimers (N = 500) have $0.809 \pm 0.007$

While the compressing bar is in motion, we acquire images of pillar deformation using a JAI/Pulnix TM-4200CL camera with a frame rate of 8 fps. Each shape requires its own unique considerations in order to perform particle tracking. For monomers, we locate all circular particles using a circular edge-finding algorithm~\cite{RieserThesis}. The displacement of the compressing bar between successive frames is about ${10^{-3}}R$, where $R$ is the large monomer radius, so linking particle tracks together is a straightforward process. For dimers, we follow monomer tracking with a procedure that effectively pairs rods together so that fluctuations in interparticle distance over time are minimized~\cite{HarringtonPRE2018}. Once a dimer is identified as a pair of rods, the positions of the rods are averaged to determine the dimer centroid position, while the difference in positions yields the dimer orientation. 

Ellipses require the most special treatment of the three particle shapes for position and orientation extraction. First, we use an edge-finding algorithm to identify points that lie at the interface between the ellipse and the bright background illumination. Second, we compute the distance map of this binary image, determining the shortest distance between each pixel and an edge pixel. This distance map contains local maxima that correspond to approximations of the ellipse centroid positions. Third, we use a Radon transform to match the local region surrounding each local maxima with that of a test image of an idealized large (small) ellipse. The purpose of this step is to determine the orientation angle, refined to a precision on the order of 0.1 deg, that provides the best match between the test and experimental images. Finally, we rotate the test image by the measured orientation in order to refine the centroid position to subpixel precision. 

The final step in particle tracking for all shapes is to suppress noise in particle positions and, when appropriate, orientations. We apply a Gaussian filter, with a time window width given by the time for the compression bar to move $\frac{2}{15}R$, to the raw measurements of horizontal position, vertical position, and orientation. The window size of the Gaussian filter becomes the effective time interval between filtered measurements of positions and orientations.

The machine learning method we use to characterize structural defects within the pillar will be introduced in full detail in Sec.~\ref{sec:SVM}. For now, we describe how we quantify particle-scale rearrangements within the granular pillar. In particular, we use the quantity $D^2_{\mathrm{min}}$ as defined in Ref.~\cite{FalkLangerPRE1998},

\begin{equation}
	\label{eq:D2min}
	D_{\mathrm{min},i}^2\left(t;\Delta t\right) = \frac{1}{n} \sum_{j=1}^n |\vec{r}_{ij}\left(t+\Delta t\right) - \textbf{E}\vec{r}_{ij}\left(t\right)|^2,
\end{equation}
where $i$ is the reference particle index, $j$ is the index of a neighboring particle, $\vec{r}_{ij}(t)$ is the relative position vector between the centroids of particles $i$ and $j$ at a time $t$, $\Delta t = 4/3~R/v_c$ is the time interval over which we measure rearrangements, $R$ is the large monomer radius, $n$ is the number of neighbors, and $\textbf{E}$ is the best fit affine transform to describe how the surrounding neighborhood of particle $i$ moves. For all shapes, $D^2_{\mathrm{min}}$ is computed based on particle centroid positions. The main distinction that needs to be made between shapes is how to determine the neighbors of a particle. For monomers, neighbors are simply surrounding particles whose centroid is within $2.5R$ of the reference monomer. For dimers, we start with each rod that makes up a reference dimer, and find surrounding rods that lie within a centroid-to-centroid distance of $2.5R$. The corresponding dimers those rods belong to, eliminating duplicates, are the dimer neighbors. For ellipses, we compute the contact distance between pairs of ellipses whose centroids are within $5.0R$ of each other, using the method of Perram and Wertheim~\cite{PERRAM1985409,FranklinShattuck2016}. This contact distance is the centroid-to-centroid distance between two ellipses, if they were brought into contact by translating one of them along their relative position vector, without rotating either ellipse. Thus, the difference between the actual centroid-to-centroid distance and the contact distance is the surface-to-surface distance. The neighbors of a reference ellipse are those whose surface-to-surface distance is within $1.0R$. 

Fig.~\ref{fig:Particles_D2min}(b) shows a snapshot of $D^2_{\mathrm{min}}$ for monomers, dimers, and ellipses. The Supplemental Material includes videos represented by these snapshots~\cite{Supplement}. For each shape, we provide a video of one full compression trial, overlaid with colors determined by $D^2_{\mathrm{min}}$. From these videos, one can immediately observe rearrangements and transient shear bands for all shapes. However, there are differences among the shapes that can be qualitatively observed in the videos. In particular, one can simply trace the boundary of the ellipse pillar and see that it must exhibit distinct rearrangement behaviors, due to the fact it is less likely to fan out like the monomer and dimer pillars. The tendency of ellipse pillars to ``barrel'' rather than expand can also be seen in Fig.~\ref{fig:Particles_D2min}(b). In Sec.~\ref{sec:SpatialCorr}, we describe how these distinct global behaviors may originate from distinct structural aspects.

\section{Support Vector Machine Construction and Performance}
\label{sec:SVM}
Given a set of structure functions that are attributed to a particle at a single time frame, we would like to classify the particle as likely, or unlikely, to rearrange over a subsequent time interval. In machine learning, this is an example of classification problems that require an initial set of data to train the model, known as the training set. In this case, the training set consists of particles known to be (non-)rearranging, in addition to associated structural features. One machine learning algorithm that can solve this type of problem is the support vector machine (SVM)~\cite{Cortes1995}. Our method additionally prescribes a continuous parameter, softness, that serves as a continuous parameter describing a particle's susceptibility to rearrangement. We start our analysis of packings made up of the three particle shapes using the established approach and workflow in Refs.~\cite{CubukSchoenholzPRL2015,SchoenholzCubukNatPhys2016,CubukSchoenholzJPCB2016,SchoenholzCubukPNAS2017,SussmanSchoenholzPNAS2017,CubukIvancicScience2017,Sharp2018,Ivancic2018}.

A monomer or dimer is considered to be rearranging if $D^2_{\mathrm{min}} > D^2_{\mathrm{min},h} = 0.25 R^2$; an ellipse rearranges if $D^2_{\mathrm{min}} > D^2_{\mathrm{min},h} = 0.175 R^2$. The high $D^2_{\mathrm{min}}$ thresholds, among the highest values measured with the parameters listed in Sec.~\ref{sec:Methods}, is chosen so that the hyperplane is trained on particles within the top $\sim1\%$ of $D^2_{\mathrm{min}}$; in other words, those with the highest degree of rearrangement. Similarly, we train non-rearranging particles with very low $D^2_{\mathrm{min}}$, specifically within the range $D^2_{\mathrm{min,n}} = 10^{-4} R^2 < D^2_{\mathrm{min}} < D^2_{\mathrm{min,l}} = 4 \times 10^{-4} R^2$. $D^2_{\mathrm{min,n}}$ represents the typical value for pillars that are stationary, setting a noise floor, and $D^2_{\mathrm{min,l}}$ is chosen to place the ceiling of non-rearrangement, such that the bottom $\sim 1\%$ of $D^2_{\mathrm{min}}$ is sampled, matching the range sampled for rearrangements. Since the pillar is driven from the top and the pillar lies on a frictional substrate, only particles near the top of the pillar actually move and are capable of rearranging. In order to ignore stationary particles, we only sample from rows within the pillar whose average speed exceeds $\frac{1}{2}v_c$, following protocol previously used in Ref.~\cite{CubukSchoenholzPRL2015}. Particles that are chosen for the training set also require a label, $y$. Rearranging particles are labeled $y = +1$, while non-rearranging particles are labeled $y = -1$.
% Also, if I train monomers with no low ceiling, performance is actually worse!

Employing the $D^2_{\mathrm{min}}$ thresholds defined above, and requiring an equal number of (non-)rearranging particles for the training set, we train on 1688 monomers collected across 10 compression trials, 2268 dimers across 20 compression trials, and 2300 ellipses across 20 compression trials. In past studies, it was shown that $\mathcal{O}(10^3)$ particles suffice for a training set that yields a hyperplane with sufficiently high accuracy in the training set, as well as similar accuracy in an untrained test set~\cite{CubukSchoenholzPRL2015}.

\begin{figure}
 	\includegraphics[width=0.5\linewidth]{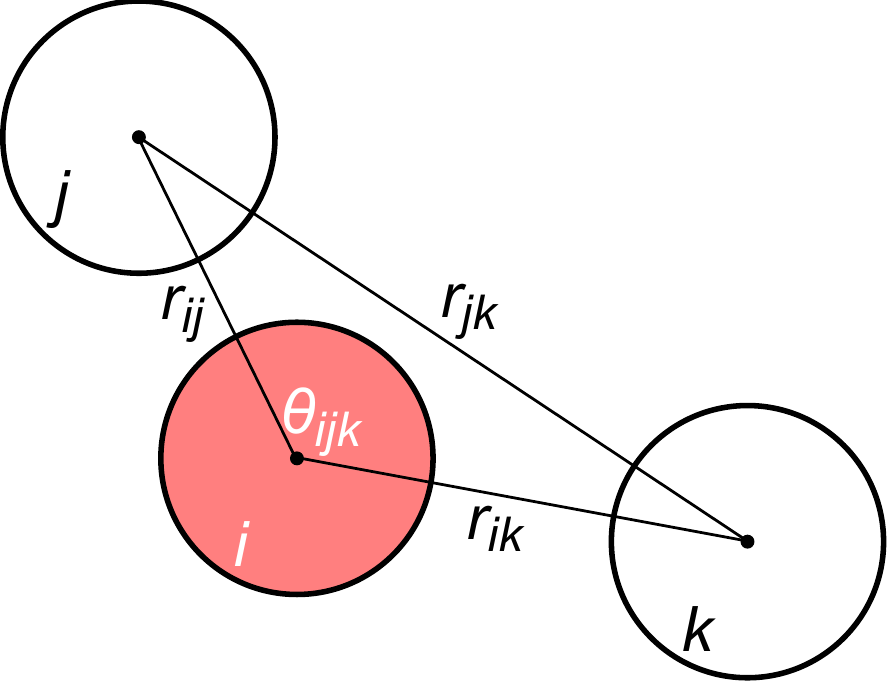}
	\caption{\label{fig:Structure_Functions} A schematic of three monomers, one reference (shaded) and two neighboring, highlighting quantities contained within the Behler-Parrinello structure functions, given in Eqs.~\ref{eq:radial} and~\ref{eq:angular_original}.} 
\end{figure}

Along with labels of whether a particle rearranges, we collect relevant descriptors quantifying the structure surrounding each particle. Following the work of Behler and Parrinello~\cite{BehlerParrinelloPRL2007}, we define two families of structure functions that highlight different structural aspects by varying parameter values. The first set of structure functions is of the form,

\begin{equation}
	\label{eq:radial}
	G_i\left(\mu\right) = \sum_j e^{-(r_{ij}-\mu)^2/\sigma^2},
\end{equation}
where $\mu$ is varied over the range $0.3 D$ to $4.9 D$ in steps of $0.1 D$, $D$ is the large monomer diameter, $\sigma = 0.1 D$ is constant, and $j$ sums over all other particles of a single species, large or small. Fig.~\ref{fig:Structure_Functions} illustrates the pairwise centroid-to-centroid distance $r_{ij}$. Given that there are two species of particles, as well as 47 different values of $\mu$, 94 features are calculated for each particle. This structure function is based purely on centriod-to-centroid distances, effectively counting the number of particles of a particular species that are a distance $\mu$ away from the reference particle. The second set of structure functions is of the form,

\begin{equation}
	\label{eq:angular_original}
	\Psi_{1i}\left(\xi,\lambda,\zeta\right) = \sum_j \sum_k e^{-(r_{ij}^2+r_{ik}^2+r_{jk}^2)/\xi^2}\left(1+\lambda\cos\theta_{ijk}\right)^\zeta,
\end{equation}
where $\xi$, $\lambda$, and $\zeta$ are varied over a range of 18 different value combinations shown in Table~\ref{table:AngularSFValues}. Fig.~\ref{fig:Structure_Functions} shows an angle $\theta_{ijk}$ measured between three particles, along with relevant interparticle distances. With three different species combinations over which to sum $j$ and $k$, this yields 54 additional features for each particle. These angular structure functions carry information regarding the size, resolution, and spatial extent of the angle between particles $i$, $j$, and $k$. Note that $1$ appears in the subscript of Eq.~\ref{eq:angular_original}, the significance of which will be specified in Section~\ref{sec:AdjFeats}. While both radial and angular structure functions are defined in terms of a sum over all other particles or combinations of particle pairs, only those within a wide interaction distance ($5.0D$ for monomers, $10.0D$ for dimers and ellipses) are included, in order to speed up computation.

\begin{table}[ht]
\centering
\begin{ruledtabular}
\caption{Parameter values for angular structure functions. $\xi$ is a length given in terms of $D$, the large monomer diameter, while $\zeta$ and $\lambda$ are dimensionless.}
\begin{tabular}{cccc}
   & $\xi/D$ & $\zeta$ & $\lambda$ \\ \hline
 1 & 2.554 & 1 & -1 \\
 2 & 2.554 & 1 & +1 \\
 3 & 2.554 & 2 & -1 \\
 4 & 2.554 & 2 & +1 \\
 5 & 1.648 & 1 & +1 \\
 6 & 1.648 & 2 & +1 \\
 7 & 1.204 & 1 & +1 \\
 8 & 1.204 & 2 & +1 \\
 9 & 1.204 & 4 & +1 \\
 10 & 1.204 & 16 & +1 \\
 11 & 0.933 & 1 & +1 \\
 12 & 0.933 & 2 & +1 \\
 13 & 0.933 & 4 & +1 \\
 14 & 0.933 & 16 & +1 \\
 15 & 0.695 & 1 & +1 \\
 16 & 0.695 & 2 & +1 \\
 17 & 0.695 & 4 & +1 \\
 18 & 0.695 & 16 & +1 
\end{tabular}
\label{table:AngularSFValues}
\end{ruledtabular}
\end{table}

Once the data in the training set has been assembled, labeled particles along with their associated structural features, we use a SVM to compute the hyperplane that best separates the two classes of data, rearranging and non-rearranging particles. Given that each particle carries 148 structural features, the computed hyperplane exists in a 148-dimensional structure space. The SVM is implemented using the {\tt scikit-learn} software package~\cite{scikit-learn}. The output from the SVM consists of a series of weights $w$, along with an intercept $b$, so that a ``softness'' parameter $S$ can be computed for subsequent data as

\begin{equation}
	\label{eq:softness_weights}
	S = \sum w\left(\mu\right)\tilde{G}\left(\mu\right) + \sum w\left(\xi,\lambda,\zeta\right)\tilde{\Psi}\left(\xi,\lambda,\zeta\right) + b,
\end{equation}
where the tildes specify that each structural feature is normalized, so that in the training set the mean values are 0 and variances are 1. The vector defined by $\left\lbrace w \right\rbrace$ represents a normal vector to the hyperplane. Therefore, $S$ can be interpreted as a signed distance from the hyperplane in structure space, with $S > 0$ corresponding to particles that are likely to undergo a rearrangement, and $S < 0$ corresponding to those that are unlikely to undergo a rearrangement. While $S$ is a purely structural quantity, comparisons with observable dynamics, e.g., $D^2_{\mathrm{min}}$, afford a protocol to assess the predictive power of this SVM method and how well it translates across data sets.

\begin{figure*}
 	\includegraphics[width=\textwidth]{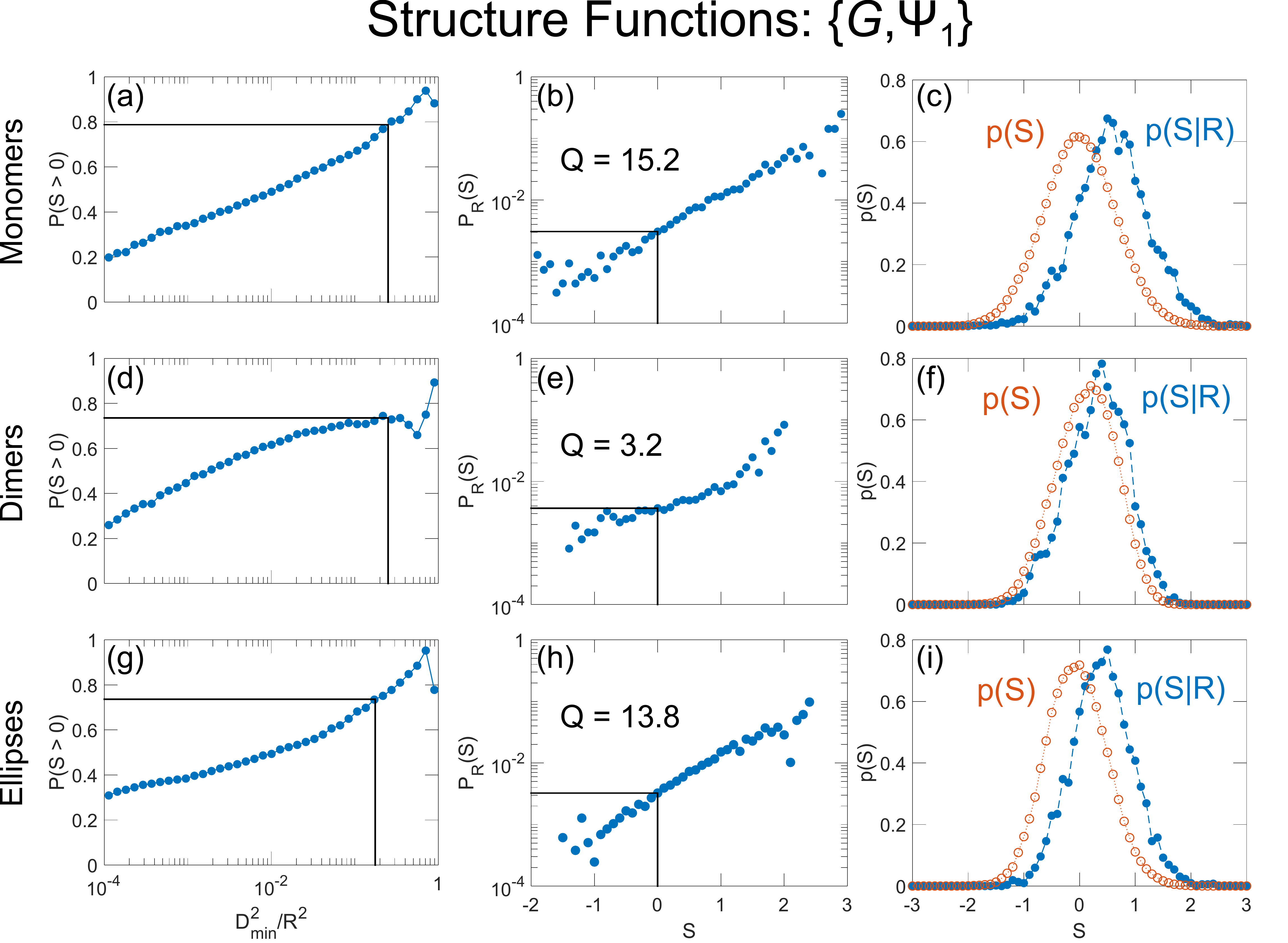}
	\caption{\label{fig:OriginalPerf_triColumn} Plots that assess the performance of softness, $S$, as computed using the Behler-Parrinello radial and angular structure functions given in Eqs.~\ref{eq:radial} and~\ref{eq:angular_original}, as an indicator of rearrangements for (a-c) monomers, (d-f) dimers, and (g-i) ellipses. The left column (a,d,g) shows the probability of a particle having positive $S$ as a function of $D^2_{\mathrm{min}}$. Vertical lines highlight $D^2_{\mathrm{min,h}}$, while horizontal lines show the corresponding probability. The middle column (b,e,h) shows the probability of a particle undergoing a rearrangement, $D^2_{\mathrm{min}} > D^2_{\mathrm{min,h}}$, as a function of softness $S$. $S = 0$ is illustrated as a vertical line, along with the corresponding probability as a horizontal line. Each plot includes $Q$, defined in Eq.~\ref{eq:Q}. The right column (c,f,i) shows probability density functions of (open circles, dotted line) $S$ of all particles and (closed circles, dashed line) $S$ belonging to particles that undergo a rearrangement.} 
\end{figure*}

Fig.~\ref{fig:OriginalPerf_triColumn} summarizes the performance of the SVM construction of softness by indicating its ability to identify particles susceptible to rearrangement in new, untrained data. Across 10 additional compression trials for packings of each shape, we highlight three plots that highlight different aspects of predictive performance. Since compression of monomers has been analyzed with softness in Ref.~\cite{CubukSchoenholzPRL2015}, we can use the monomer plots as a baseline to compare with other shapes.
% Sentence before last: We also assign $S$ only to particles that lie within a row in the pillar with average speed greater than $\frac{1}{2}v_c$, in order to only consider particles that are activated by the compressing bar.

First, we consider the probability of a particle having $S > 0$ as a function of its $D^2_{\mathrm{min}}$, shown in Fig.~\ref{fig:OriginalPerf_triColumn}(a,d,g). In this plot, we check to see if there is a generally monotonic relationship, ideally with the upward trend growing stronger for higher $D^2_{\mathrm{min}}$. All three shapes exhibit a positive correlation between these quantities, although dimers seem to reach a capacity for high values of $D^2_{\mathrm{min}}$.

Second, we plot the probability that a particle rearranges, $D^2_{\mathrm{min}} > D^2_{\mathrm{min,h}}$, as a function of its $S$ in Fig.~\ref{fig:OriginalPerf_triColumn}(b,e,h). Again, we are looking for a monotonic relationship between these quantities, ideally one that spans multiple orders of magnitude in probability. This aspect is also summarized in the metric $Q$ originally defined in Ref.~\cite{CubukSchoenholzJPCB2016},

\begin{equation}
	\label{eq:Q}
	Q = \frac{P_R\left(S>\mu+\sigma\right)}{P_R\left(S<\mu-\sigma\right)},
\end{equation}
where $\mu$ and $\sigma$ are the mean and standard deviation of $S$ for all particles, and $Q$ is the ratio of the rearrangement probability of particles with high $S$ values to that of particles with low $S$ values. A larger $Q$ value indicates that the structural metric is strongly correlated with the likelihood to rearrange. In terms of classification performance, one can interpret $Q$ as the ratio of the percentage of particles with high $S$ values that are ``true positives'' to the percentage of particles with low $S$ values that are ``false negatives.'' In Fig.~\ref{fig:OriginalPerf_triColumn}, we see that monomers and ellipses have similar $Q$ values, while that of dimers is an order of magnitude lower. This observation is also reflected in the $P_R$ vs. $S$ plot for dimers appearing significantly flatter.

Third, perhaps most telling, we compare the probability density distributions of $S$ of all particles, with that of particles that are known to rearrange with $D^2_{\mathrm{min}} > D^2_{\mathrm{min,h}}$. For this comparison, as in Ref.~\cite{SchoenholzCubukNatPhys2016}, we expect that the distribution of $S$ for rearranging particles to be shifted to the right, so that rearranging particles tend to have $S > 0$. These plots are shown in Fig.~\ref{fig:OriginalPerf_triColumn}(c,f,i). We certainly see a clear separation in distributions in the case of monomers, in which 82\% of rearranging particles have $S > 0$, compared with the overall distribution of $S$ that is centered near $S = 0$. A similar separation is apparent in ellipses, but those of dimers are much less separated, to the extent that they essentially lie right on top of each other. This indicates that a dimer, regardless of whether it rearranges, would exhibit similar ranges in $S$. 

While successful for monomers and ellipses, the collective results in Fig.~\ref{fig:OriginalPerf_triColumn}(d,e,f) represent a clear failure of $S$ to provide a meaningful indicator of rearrangement susceptibility for dimers. For elongated shapes in general, this could be expected. All of the structure functions, as presently defined, only rely on centroid-to-centroid distances and angles. The relative orientations of dimers do not enter anywhere into Eqs.~\ref{eq:radial} and~\ref{eq:angular_original}, so perhaps the SVM is not given enough relevant features regarding local structure. While ellipses exhibit similar performance as monomers, it stands to reason that attempts to improve the training performance of dimers may bear some relevance for ellipses, as both are elongated shapes with attributable orientations and local instances of alignment. In Sec.~\ref{sec:AdjFeats}, we consider strategies to improve the utility of $S$ in packings of dimers, and their effects on ellipses.

% Current Q values from training done with angular features \xi <= 2.554

% Q(monomers -- noLongAngs MAY2018) = 15.1802 ; mean of all softness values = -7.5e-4 ; st. dev. of all softness values = 0.65 ; mean of rearrangement softness = 0.58 ; st. dev. of rearrangement softness = 0.65 ; 82% of R's have S > 0 ; self-score (C = 1) = 0.84

% Q(dimer centriods -- monomerFeats) = 3.1755 ; mean of all softness values = 0.11 ; st. dev. of all softness values = 0.54 ; mean of rearrangement softness = 0.32 ; st. dev. of rearrangement softness = 0.54 ; 72% of R's have S > 0

% Q(ellipse centroids -- monomerFeats) = 13.8402 ; mean of all softness values = -0.046 ; st. dev. of all softness values = 0.56 ; mean of rearrangement softness = 0.41 ; st. dev. of rearrangement softness = 0.57 ; 77% of R's have S > 0 

\section{Adjusted Structure Functions for Elongated Particles}
\label{sec:AdjFeats}
% Q(dimer parts -- noLongAngs-dimerParts) = 18.9817 ; mean of all softness values =  -0.057 ; st. dev. of all softness values = 0.60 ; mean of rearrangement softness =  0.55 ; st. dev. of rearrangement softness = 0.60 ; 82% of R's have S > 0 ; self score (C = 0.1) = 0.80

As previously noted, dimers are pairs of monomers that are bonded together. In fact, the first steps in analyzing compression trials with dimers is to track their constituent rods as if they are discrete monomers. Therefore, a simple way to ``improve'' softness training with dimers is to treat each particle as two distinct monomers. Under this simple approach, the resulting performance benchmarks closely resemble those in Fig.~\ref{fig:OriginalPerf_triColumn}(a,b,c). Of course, this would not be a useful procedure to extend to ellipses since they are continuous shapes and do not easily segment into monomer components. Furthermore, this approach alone still does not address how orientational structure can be explicitly accounted for in determining a generically elongated particle's likelihood to rearrange. It does, however, provide a useful starting point. 

\begin{figure}
 	\includegraphics[width=0.8\linewidth]{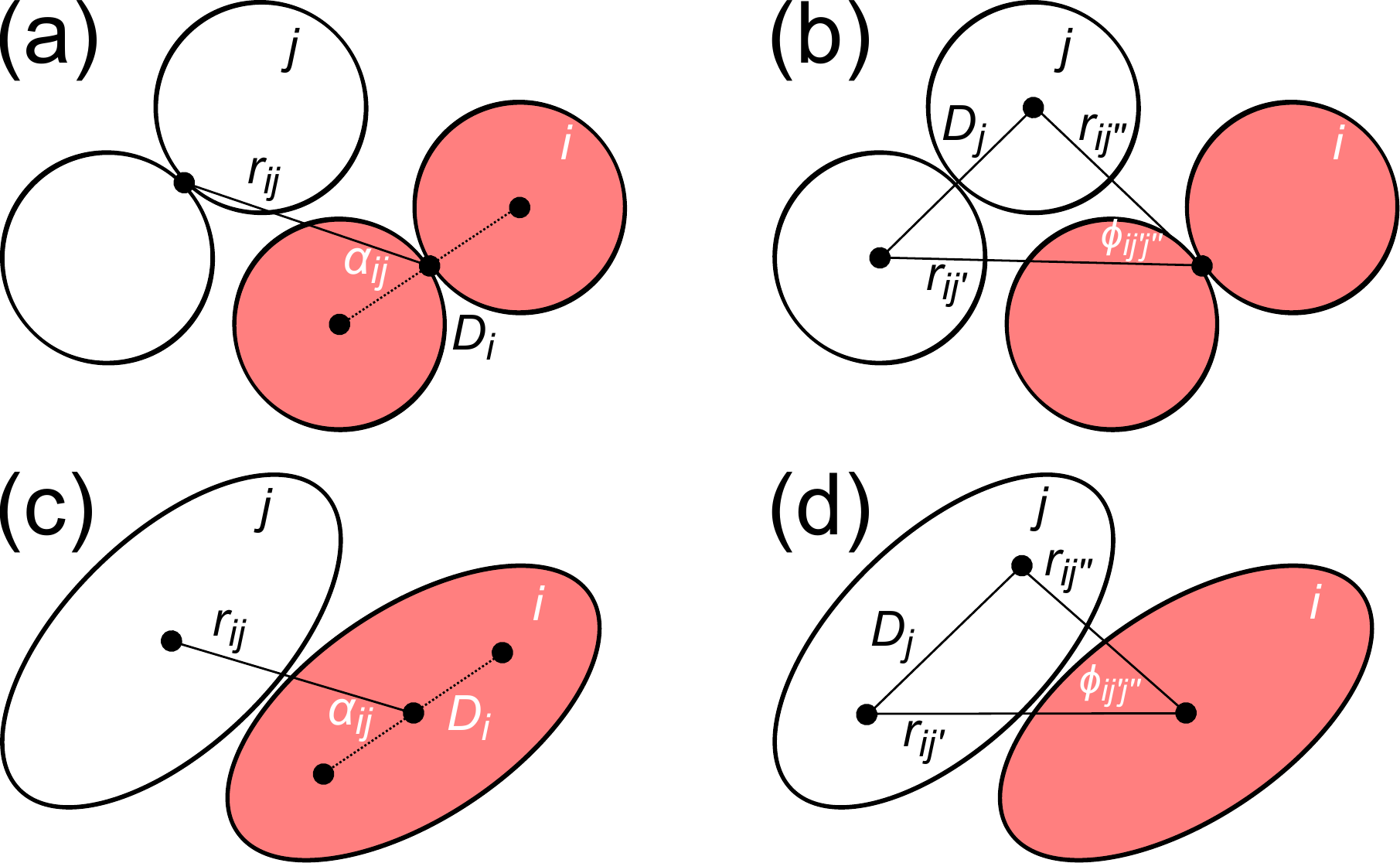}
	\caption{\label{fig:Elongated_Structure_Functions} Schematics illustrating the definitions of orientational structure functions, given in Eqs.~\ref{eq:angular_2} and~\ref{eq:angular_3}. The left column (a,c) corresponds to $\Psi_2$, which is determined by the relative angle between the orientation of the reference particle (shaded) and the position vector to another particle. The right column (b,d) corresponds to $\Psi_3$, which includes a fan angle that is swept out along the orientation of a neighboring particle.} 
\end{figure}

When the dimers are trained as if they are discrete monomers, there are three types of angular groups that are summed over. There is one type in which the 3 monomers lie on different dimers, one type in which the reference monomer shares a dimer with one of the other monomers, and a final type in which the monomers that are not the reference monomer make up a dimer themselves. How would these structure functions translate if they are ascribed to the overall dimers? This question is motivated by the general tendency that if a constitutive monomer rearranges, so does the dimer it constitutes. Also, this would allow for structural characterizations that are explicitly dimer-based, rather than constitutive monomer-based, easing their application to ellipses and other elongated shapes. Fig.~\ref{fig:Elongated_Structure_Functions} illustrates how the latter two angular groups are captured for dimers, then extended to ellipses. The first angular group is already accounted for in $\Psi_1$ and Fig.~\ref{fig:Structure_Functions}, which only concerns centroid-to-centroid distances and angles.

Fig.~\ref{fig:Elongated_Structure_Functions}(a) shows how the second angular group could translate to dimers. In this case, the relevant angle is that between the orientation of the reference dimer and the position vector to another dimer centroid. Following the original Behler-Parrinello definitions, this new structure function is given by 

\begin{equation}
	\label{eq:angular_2}
	\Psi_{2i}\left(\xi,\lambda,\zeta\right) = \sum_j e^{-(2r_{ij}^2+D_{i}^2)/\xi^2}\left(1+\lambda\cos\alpha_{ij}\right)^\zeta,
\end{equation}
where $D_i$ is the distance between constitutive monomers of $i$ and $\alpha_{ij}$ is the angle between the orientation of $i$ and the position vector from $i$ to $j$. Two supplemental angles can be defined this way, so, for consistency, we always take $\alpha_{ij}$ to be the smaller angle. To better understand the physical implications of this structure function, consider only this hypothetical pair of dimers: the reference dimer $i$ and one other dimer $j$. In other words, suppose the ensemble is made up of 2 dimers. One can imagine tuning the value of $\Psi_{2i}$ by simply rotating the orientation of reference dimer $i$. When $\lambda = +1$, the value of $\Psi_{2i}$ will increase when $\alpha_{ij}$ is small. Physically, this indicates the orientation of the major axis the reference dimer closely follows the position vector to another dimer. Likewise, when $\lambda = -1$, the value of $\Psi_{2i}$ increases as $\alpha_{ij}$ approaches $\frac{\pi}{2}$ rad ($90 ^\circ$), where the minor axis aligns with the relative position vector. 

The third angular group translated to dimers is illustrated in Fig.~\ref{fig:Elongated_Structure_Functions}(b), in which the angle is swept out along the orientation of the neighboring dimer. Following again the original Behler-Parrinello definition, this new structure function is given by

\begin{equation}
	\label{eq:angular_3}
	\Psi_{3i}\left(\xi,\lambda,\zeta\right) = \sum_j e^{-(r_{ij'}^2+r_{ij''}^2+D_{j}^2)/\xi^2}\left(1+\lambda\cos\phi_{ij'j''}\right)^\zeta,
\end{equation}
where $D_j$ is the distance between constitutive monomers of $j$ and $\phi_{ij'j''}$ is the angle between the centroid of $i$ and the constitutive monomers of $j$. We can perform a similar thought experiment as in the previous paragraph to elucidate the physical aspects captured by $\Psi_{3i}$. Consider the reference dimer $i$ and only one other dimer $j$. The value of $\Psi_{3i}$ will now change as dimer $j$, not reference dimer $i$, is rotated. When $\lambda = +1$, $\Psi_{3i}$ grows when $\phi_{ij'j''}$ is small. This corresponds to the orientation of the major axis of $j$ mostly aligning with the relative position vector from $i$ to $j$. When $\lambda = -1$, $\Psi_{3i}$ increases when $\phi_{ij'j''}$ is large, as in the minor axis of $j$ aligning with the relative position vector. Moreover, the correlation of $\Psi_{2i}$ and $\Psi_{3i}$ is also telling. If $\Psi_{2i}$ and $\Psi_{3i}$ are both large for a single pair of dimers, then they exhibit orientational alignment either along ($\lambda = +1$) or normal to ($\lambda = -1$) their relative position vector. If $\Psi_{2i}$ is large while $\Psi_{3i}$ is small, or vice versa, then the two dimers exhibit some form of antialignment.

Fig.~\ref{fig:Elongated_Structure_Functions}(c,d) shows how $\Psi_2$ and $\Psi_3$ would correspond to ellipses. In this case, we take $D_i$ and $D_j$ to be twice the minor axis of $i$ and $j$, respectively. As with dimers, $\Psi_2$ depends on the orientation of the reference ellipse, while $\Psi_3$ depends on the orientations of surrounding ellipses. The combination of both provide explicit orientational and alignment considerations related to structure. Note that for monomers, $\Psi_2$ and $\Psi_3$ are ill-defined due to their symmetric shape, so this adjustment does not affect previous calculations for monomers. 

% Q(dimers adjusted -- noLongAngs) = 17.1855 ; mean of all softness values = -0.011 ; st. dev. of all softness values = 0.59 ; mean of rearrangement softness = 0.54 ; st. dev. of rearrangement softness = 0.62 ; 80% of rearr.'s have S > 0 ; self score (C = 1) = 0.83

% Q(ellipses adjusted -- 0p175) = 56.6773 ; mean of all softness values = -0.12 ; st. dev. of all softness values = 0.7083 ; mean of rearrangement softness = 0.7384 ; st. dev. of rearrangement softness = 0.7468 ; 85% of rearr.'s have S > 0 ; self score (C = 1) = 0.83

\begin{figure*}
 	\includegraphics[width=\textwidth]{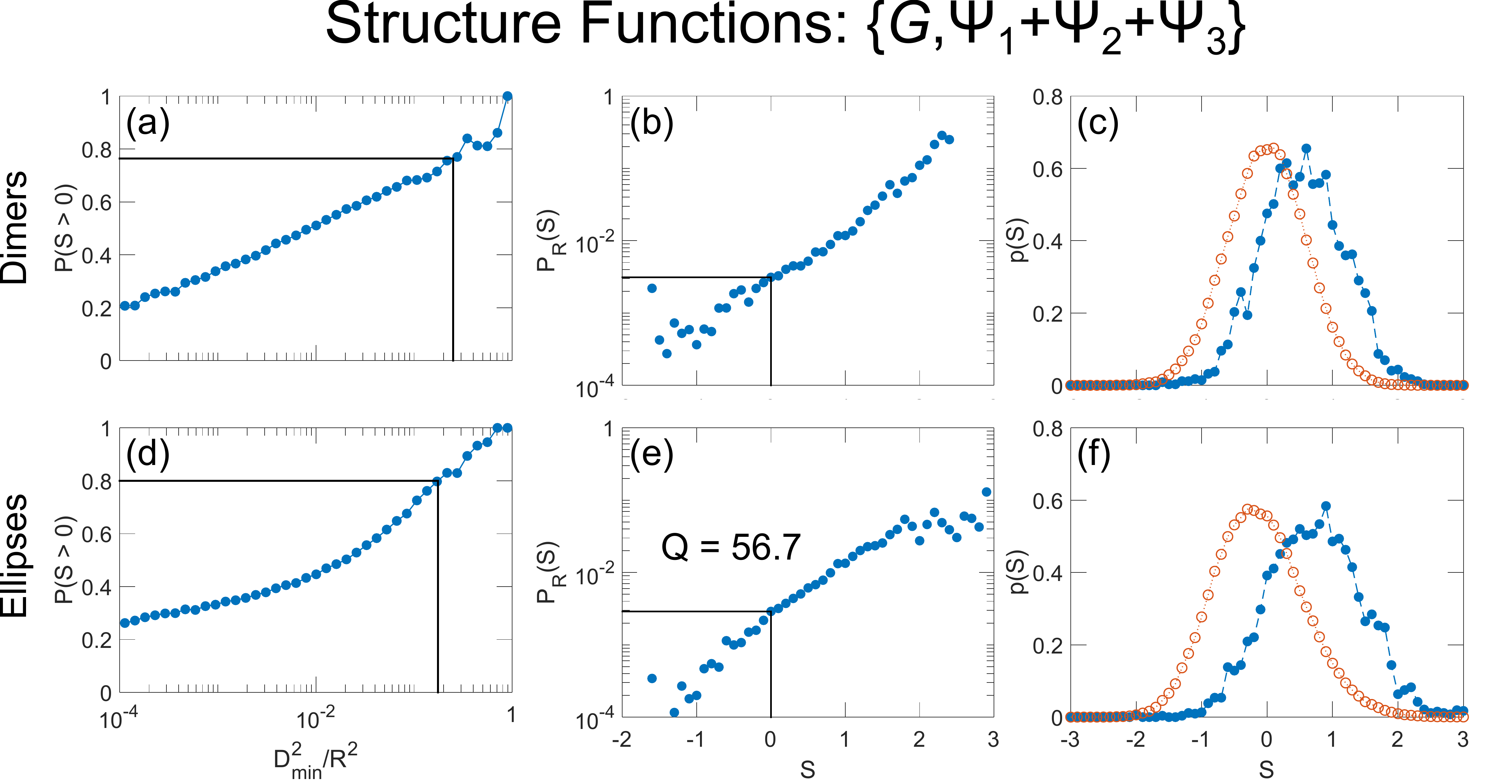}
	\caption{\label{fig:elongatedFixedPerformance_combined} Plots that assess the performance of $S$ in dimers and ellipses when computed with a set of structure functions comprised of radial (Eq.~\ref{eq:radial}) functions and the sum of angular (Eq.~\ref{eq:angular_original}) and orientational (Eqs.~\ref{eq:angular_2} and~\ref{eq:angular_3}) functions. The definition of each plot is given in Fig.~\ref{fig:OriginalPerf_triColumn}.}
\end{figure*}

If all three types of angular groups are combined, given a single row of $\xi$, $\lambda$, and $\zeta$ values from Table~\ref{table:AngularSFValues}---as they would in the case of constitutive monomer training---the full angular structure function for a dimer or ellipse is $\Psi = \Psi_1 + \Psi_2 + \Psi_3$. This preserves the same number of features given to the SVM in Fig.~\ref{fig:OriginalPerf_triColumn}. In Fig.~\ref{fig:elongatedFixedPerformance_combined}, we show performance metrics under these newly defined structure functions. Indeed, we now see performance in dimers that matches that of monomers, especially in terms of well-separated $S$ distributions and a more reasonable $Q$ value. The performance of ellipses is, in fact, even better with these modified structure functions.

% Q(dimers separate -- noLongAngs) = 21.1566 ; mean of all softness values = -0.0442 ; st. dev. of all softness values = 0.7094 ; mean of rearrangement softness = 0.6776 ; st. dev. of rearrangement softness = 0.7689 ; 82% of rearr.'s have S > 0 ; self score (C = 1) = 0.83

% Q(ellipses separate -- 0p175) = 51.5642 ; mean of all softness values = -0.1456 ; st. dev. of all softness values = 0.7578 ; mean of rearrangement softness = 0.7911 ; st. dev. of rearrangement softness = 0.7856 ; 85% of rearr.'s have S > 0 ; self score (C = 1) = 0.84

\begin{figure*}
 	\includegraphics[width=\textwidth]{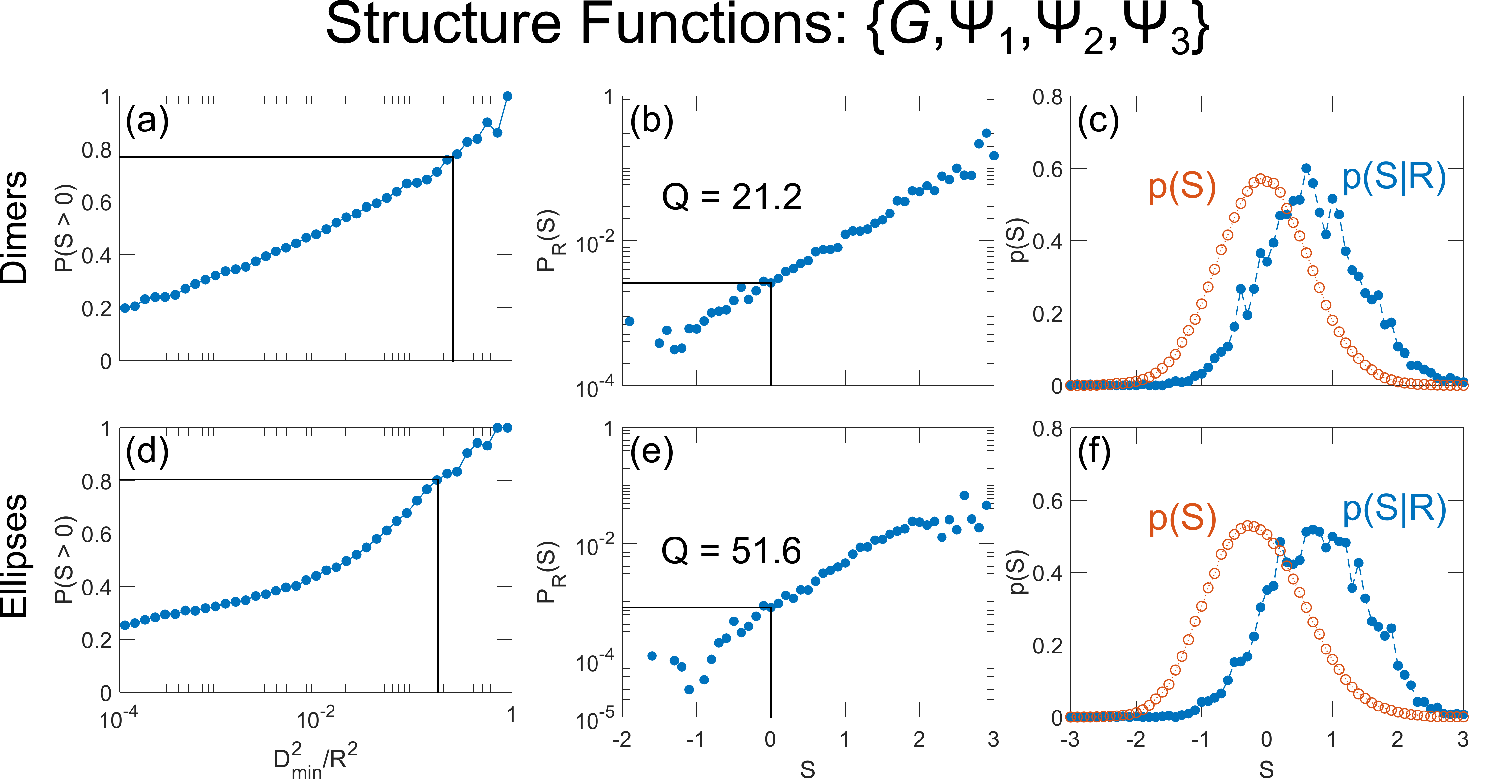}
	\caption{\label{fig:elongatedFixedPerformance_separate} Plots that assess the performance of $S$ in dimers and ellipses when computed with a set of structure functions comprised of radial (Eq.~\ref{eq:radial}), angular (Eq.~\ref{eq:angular_original}), and orientational (Eqs.~\ref{eq:angular_2} and~\ref{eq:angular_3}) functions. As opposed to Fig.~\ref{fig:elongatedFixedPerformance_combined}, angular and orientational structure functions are given to the SVM as distinct features. The definition of each plot is given in Fig.~\ref{fig:OriginalPerf_triColumn}.}
\end{figure*}

As will become apparent in Section~\ref{sec:RFE}, it can be difficult to assess the structural significance of the combined angular structure functions. When $\Psi_1$, $\Psi_2$, and $\Psi_3$ are added together, different structural considerations are added together, so their unique significance is generally lost. For this reason, we also perform calculations of $S$ with the angular and orientational structure functions treated as separate features in the SVM. When doing so, this increases the number of SVM features for dimers and ellipses from $148$ to $220$. The results in Fig.~\ref{fig:elongatedFixedPerformance_separate} are practically unaffected in terms of overall predictive performance. For the purpose of interpreting the importance of individual structural aspects in Section~\ref{sec:RFE}, such that we can separate radial, angular, and orientational effects, the remaining results in the article treat $\Psi_1$, $\Psi_2$, and $\Psi_3$ as separate features. 

\section{Spatial Correlations of Rearrangements and Softness}
\label{sec:SpatialCorr}

Up to this point, we have mainly reported on how to improve a previously-established SVM approach to generate a particle-based parameter that is indicative of rearrangements in disordered packings of elongated particles. We have not yet discussed the physical implications of the softness field, much less how it relates to the rearrangement field given by $D^2_{\mathrm{min}}$. In this section, we begin gleaning physical meaning from our results by considering the spatial extent of the rearrangement and softness fields. To reiterate, for dimers and ellipses, we consider calculations of $S$ obtained from separate structure functions, $\lbrace G, \Psi_1, \Psi_2, \Psi_3\rbrace$, as illustrated in Fig.~\ref{fig:elongatedFixedPerformance_separate}.

\begin{figure}
 	\includegraphics[width=\linewidth]{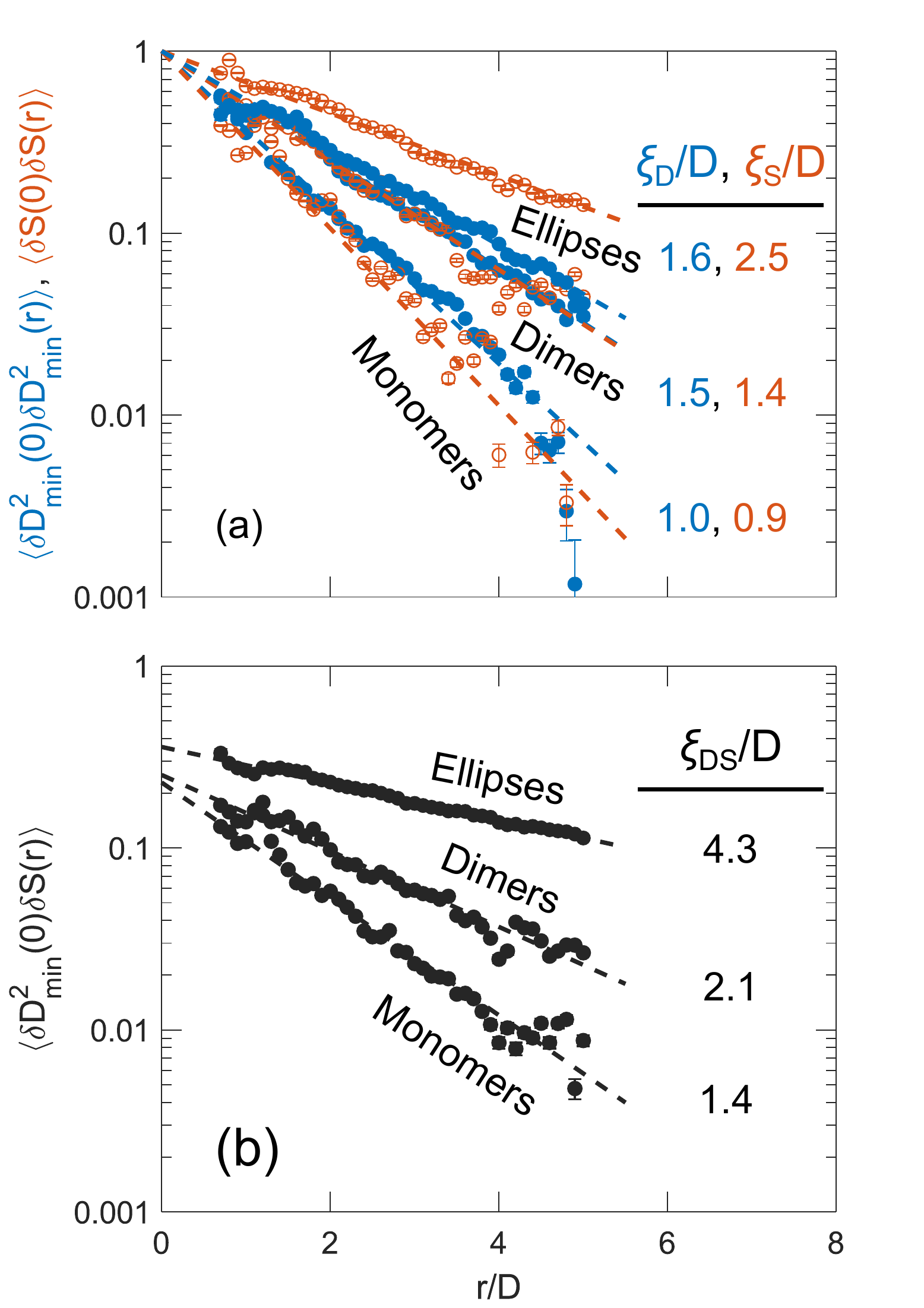}
	\caption{\label{fig:spatialAutoCorrs_Flicker_separate} (a) Spatial autocorrelations of rearrangements ($D^2_{\mathrm{min}}$, closed circles) and softness ($S$, open circles) for monomers, dimers, and ellipses. The dashed lines are fits to the function $e^{-r/\xi}$, where $\xi$ is the correlation length of either rearrangements, $D^2_{\mathrm{min}}$ (D), or softness, $S$ (S). (b) Spatial cross-correlations of rearrangements ($D^2_{\mathrm{min}}$) and softness ($S$) for monomers, dimers, and ellipses. The dashed lines are fits to the function $Ce^{-r/\xi_{DS}}$, where $\xi_{DS}$ is the cross-correlation length of rearrangements and softness. In both plots, lengths are quantified in terms of $D$, the large monomer diameter.}
\end{figure}

We quantify the spatial extent of $D^2_{\mathrm{min}}$ and $S$ by computing their spatial autocorrelations. For a single field, say $D^2_{\mathrm{min}}$, we examine the length over which $\langle\delta D^2_{\mathrm{min}}(0)\delta D^2_{\mathrm{min}}(r)\rangle$ decays. Specifically, for a single particle,

\begin{equation}
	\label{eq:norm_fluct}
	\delta D^2_{\mathrm{min}} = \frac{D^2_{\mathrm{min}}-\langle D^2_{\mathrm{min}}\rangle}{\sqrt{\langle\left(D^2_{\mathrm{min}} - \langle D^2_{\mathrm{min}}\rangle\right)^2\rangle}}
\end{equation}
is the normalized fluctuation of $D^2_{\mathrm{min}}$ at a particular time frame. $\delta D^2_{\mathrm{min}}(0)$ refers to the normalized fluctuation of a reference particle, while $\delta D^2_{\mathrm{min}}(r)$ is that of another particle at a distance $r$ away, in terms of centroid-to-centroid distance. We average the product of normalized fluctuations, $\delta D^2_{\mathrm{min}}(0)\delta D^2_{\mathrm{min}}(r)$, over all unique particle pairs. This procedure is repeated for the $S$ field. Fig.~\ref{fig:spatialAutoCorrs_Flicker_separate}(a) shows that for all shapes, the autocorrelations of both $D^2_{\mathrm{min}}$ and $S$ decay as $e^{-r/\xi}$, such that $\xi_D$ can be defined as the correlation length of rearrangements ($D^2_{\mathrm{min}}$), while $\xi_S$ is the correlation length of defects or softness ($S$). Note that all time frames are used for these calculations, as there is no discernible trend in the evolution of $\xi_D$ or $\xi_S$ as the pillar is compressed over time.
% If it needs to be mentioned, we shift the correlations by an empirically determined vertical shift, so that we are highlighting just the rate of decay

Reinforcing the relationship between rearrangements and defects, $\xi_D$ and $\xi_S$ are very similar for monomers and dimers. This common exponential relationship was also observed in Ref.~\cite{CubukIvancicScience2017}, for disordered solids with symmetric constitutive particles spanning 8 decades in size. Interestingly, ellipses exhibit a larger $\xi_S$ compared to $\xi_D$. As expected, every shape undergoes highly localized rearrangements, with correlations on the order of individual particle sizes. While dimers and ellipses exhibit a larger $\xi_D$ than monomers, they fall within the broad spectrum of nearest neighbor centroid-to-centroid distances. Likewise for monomers and dimers, defects exhibit similar localization. One way to interpret this is to consider the full distillation of structure significantly changing from one particle to the next. For ellipses, $\xi_S$ extends well beyond nearest neighbor distances, indicating that overall structure is less spatially variant. This could be a result of the elliptical shapes encouraging local ordering, whereas the circular surfaces of dimers do not directly inform the orientation, so natural ordering is suppressed.

We also consider the spatial cross-correlation of $D^2_{\mathrm{min}}$ and $S$ using the same normalized fluctuations given in Eq.~\ref{eq:norm_fluct}, and observing the decay of $\langle\delta D^2_{\mathrm{min}}(0)\delta S(r)\rangle$. In Figure~\ref{fig:spatialAutoCorrs_Flicker_separate}(b), we still observe an exponential decay of the form $Ce^{-r/\xi_{DS}}$, where $\xi_{DS}$ quantifies the correlation length between $D^2_{\mathrm{min}}$ and $S$. Again, this helps reinforce the positive correlation between individual measurements of $D^2_{\mathrm{min}}$ and $S$, and indicates their relationship decays over a characteristic length. Comparing the autocorrelation and cross-correlation lengths, we see that $\xi_{DS}$ is slightly larger, suggesting that there may be feedback response between $S$ and $D^2_{\mathrm{min}}$ that covers a wider distance. Still, $\xi_{DS}$ is indicative of localized behavior for monomers and dimers. Ellipses, meanwhile, exhibit a relationship between $D^2_{\mathrm{min}}$ and $S$ that extends well beyond individual particle sizes.

Examining the spatial extent of the rearrangement and softness fields are useful in elucidating their predictive relationship. In other words, they address whether the characteristic size of a rearrangement corresponds with that of a rearrangement-susceptible region. Note that this is a comparison between a static quantity, $S$, with a dynamic quantity that is measured over a subsequent time interval, $D^2_{\mathrm{min}}$. At this stage, we are also interested in how overall structure changes over the entire time interval of a rearrangement. 

\begin{figure*}
 	\includegraphics[width=\textwidth]{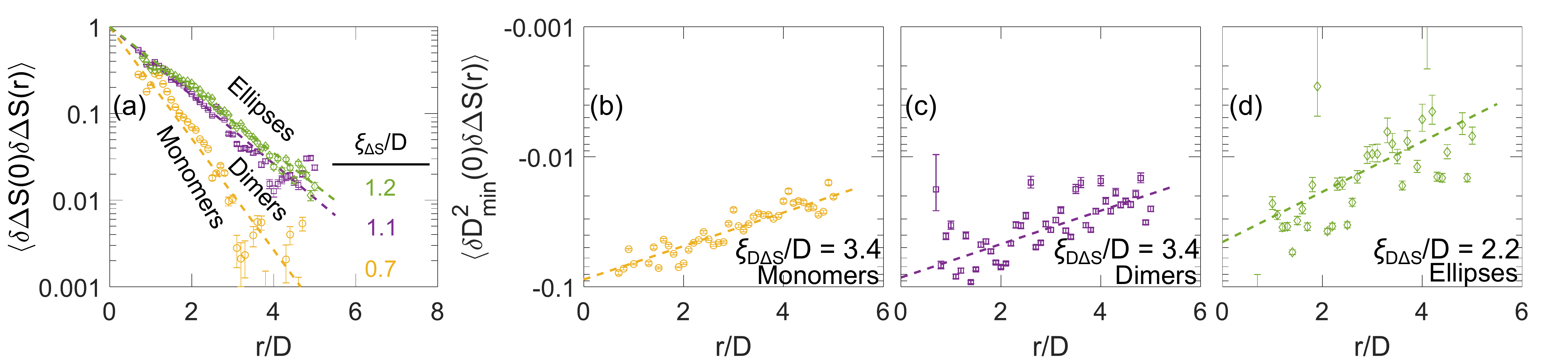}
	\caption{\label{fig:diffSoftness} (a) Spatial autocorrelations of the change in softness ($\Delta S$), over the same interval used to compute $D^2_{\mathrm{min}}$, for monomers (yellow circles), dimers (purple squares), and ellipses (green diamonds). The dashed lines are fits to the function $e^{-r/\xi_{\Delta S}}$, where $\xi_{\Delta S}$ is the correlation length of changes in softness. (b)-(d) Spatial cross-correlations of rearrangements ($D^2_{\mathrm{min}}$) and change in softness ($\Delta S$) for (b) monomers, (c) dimers, and (d) ellipses. The dashed lines are fits to the function $Ce^{-r/\xi_{D\Delta S}}$, where $\xi_{D\Delta S}$ is the cross-correlation length of rearrangements and softness. In all plots, lengths are quantified in terms of $D$, the large monomer diameter.} 
\end{figure*}

To quantify this, we examine the change in softness, $\Delta S$, for each particle over the same time interval used to compute $D^2_{\mathrm{min}}$. Figure~\ref{fig:diffSoftness}(a) shows the spatial autocorrelation of $\Delta S$ for all three shapes. Again, an exponential relation of the form $e^{-r/\xi_{\Delta S}}$ is apparent, where $\xi_{\Delta S}$ is the correlation length of changes in softness. We find correlation lengths that are highly localized, in line with the spatial extent of rearrangements. This alleviates the discrepancy in $\xi_D$ and $\xi_S$ previously observed in ellipses. While their static structure varies over longer length scales, dynamic changes to structure remain highly localized.

To further consider the relationship between rearrangements and structural changes, we examine the spatial cross-correlation between $D^2_{\mathrm{min}}$ and $\Delta S$, shown in Fig.~\ref{fig:diffSoftness}(b)-(d). For each shape, we tend to observe a negative correlation between the quantities. In other words, a rearranging particle with high $D^2_{\mathrm{min}}$ has a general tendency to reduce its own softness, plus that of surrounding particles, by the end of the rearrangement. Interestingly, the spatial cross-correlation between $D^2_{\mathrm{min}}$ and $\Delta S$ retains an exponential relationship for every shape. Furthermore, the spatial extents of their correlation, $\xi_{D\Delta S}$, are larger than the generally localized effects described above. In fact, ellipses, while exhibiting a fair amount of noise in the cross-correlation measurements, seem to undergo rearrangements that affect structure over the smallest spatial range. Meanwhile, monomers and dimers rearrange in a way that affects structure over a significantly larger range.

The collective results regarding correlation lengths could actually inform qualitative observations from the videos of $D^2_{\mathrm{min}}$-colored particles in the Supplemental Material~\cite{Supplement}. For every shape, one can observe that rearrangements---spikes in $D^2_{\mathrm{min}}$ values---tend to be restricted to individual particles. Transient shear bands, lines of high $D^2_{\mathrm{min}}$, naturally form and dissipate over time. From Fig.~\ref{fig:spatialAutoCorrs_Flicker_separate}(a), regions of monomers and dimers that are susceptible to rearrangement remain similarly localized over time. Meanwhile, larger clusters of ellipses are equally likely to rearrange. However, the changes in rearrangement susceptibility, a proxy for overall structure, is localized for all shapes, as shown in Fig.~\ref{fig:diffSoftness}(a). These aspects help explain the unique ``barreling'' behavior of ellipses described in Sec.~\ref{sec:Methods}. The localization of rearrangements, structural defects, and structural changes for monomers and dimers allow the top of the pillar to readily slip outwards, yielding an overall deformation shape that is similar for both shapes. In contrast, structurally similar ellipses are more clustered, while the interface of rearrangement and structural changes remain particle-sized. Thus, these characteristics can manifest in the form of clusters of ellipses that collectively compress over each other, or ``barrel.''

\section{Structure Function Importance from Recursive Feature Elimination}
\label{sec:RFE}

Another physical interpretation of softness we would like to address is how one could identify structural aspects that are the most crucial in determining the softness of a particle. Our analytical approach so far involves taking a large volume of structural measurements and distilling all of that information into a single parameter that correlates with the susceptibility to rearrange. We would now like to attempt to answer the inverse question. Given a hyperplane that best separates rearranging and non-rearranging particles, which individual structural aspects are the most discriminatory? Following this line of reasoning opens avenues to explore determinations of an optimal subset of structure functions that balances accuracy and computational effort. For dimers and ellipses, we restrict our discussion to calculations of softness using separate structure functions, $\lbrace G, \Psi_1, \Psi_2, \Psi_3\rbrace$, as shown in Fig.~\ref{fig:elongatedFixedPerformance_separate}.

The hyperplane already has a normal vector with components that correspond to the weights $w$ assigned to each structure function, as shown in Eq.~\ref{eq:softness_weights}. As a quick assessment of feature importance, one could simply rank the features based on the magnitudes of $w$. However, this is unlikely to be the best way of assessing feature importance. As constructed, all of the structures functions, radial, angular, and orientational, can exhibit large degrees of covariance. For example, the presence of a large monomer neighbor at $1.0D$ away from another reduces the ability to place one at $1.2D$. As a result, the full set of structure functions fails to satisfy orthogonality, so $\left\lbrace w \right\rbrace$ should not be interpreted as wholly independent components. Evaluating feature importance in the presence of covariant features, as well as feature selection that relies on regularization are major branches of ongoing work in the machine learning community~\cite{Peng2005,FIRM}. 

\begin{figure}
 	\includegraphics[width=0.85\linewidth]{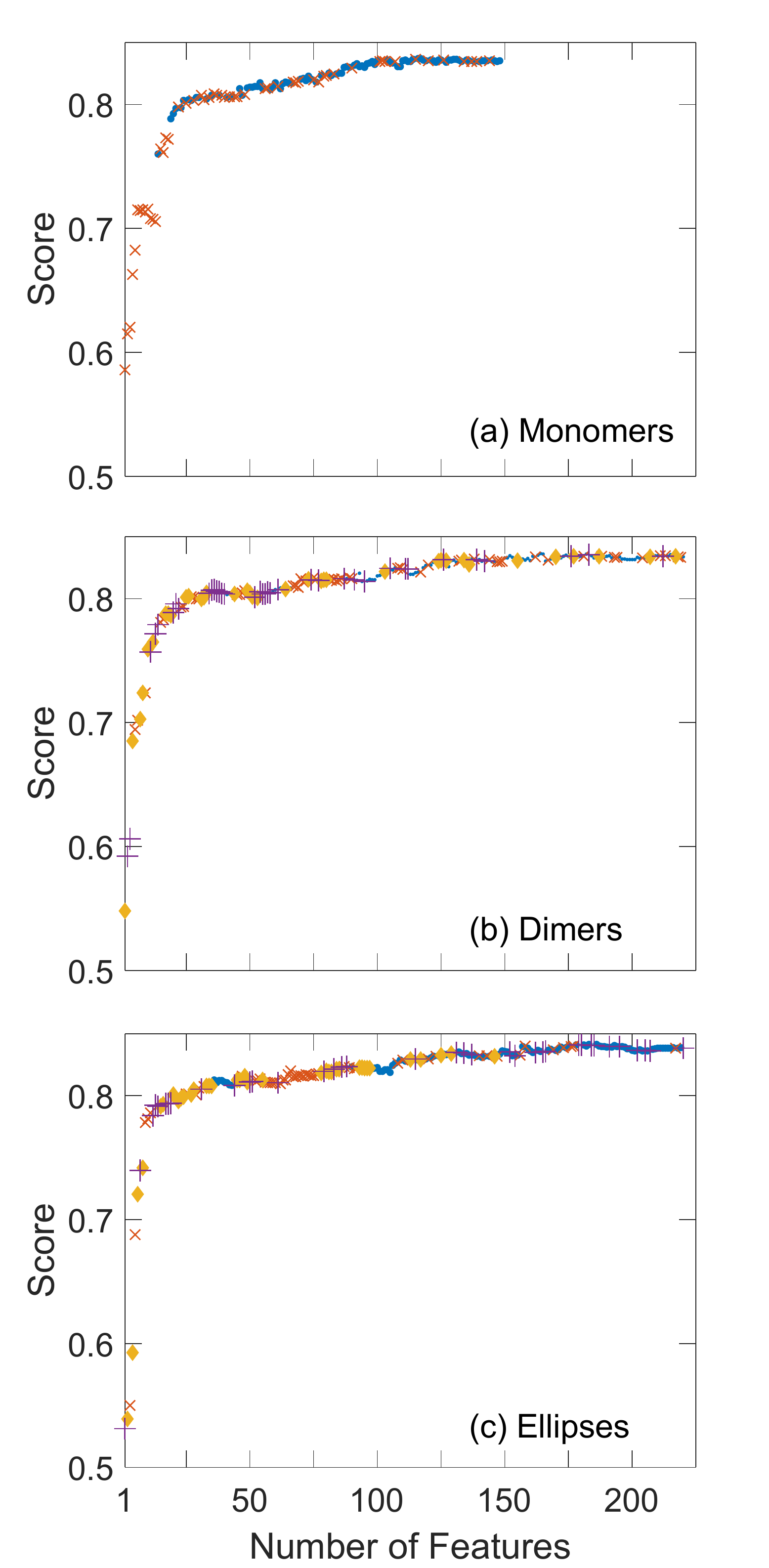}
	\caption{\label{fig:RFE_scores_separateAngular} Mean training accuracy, or ``score,'' as a function of the number of features used during recursive feature elimination (RFE) for (a) monomers, (b) dimers, and (c) ellipses. Score specifies the fraction of training examples that fall on the proper side of the calculated hyperplane. The symbols at each step correspond to the feature type that is removed on the next recursion: $G$ ($\bullet$), $\Psi_1$ ($\times$), $\Psi_2$ ($\blacklozenge$), and $\Psi_3$ ($+$).}
\end{figure}

An established method of reducing dimensionality to a subset of important features, especially in the context of SVM, is recursive feature elimination (RFE)~\cite{Guyon2002}. RFE starts with the full set of weights $w$. The feature with the smallest $w$ in magnitude is thrown out of the SVM model. The hyperplane is recalculated with the remaining features, and this process repeats until only one feature remains. To track the impact of incrementally removing features in this fashion, Fig.~\ref{fig:RFE_scores_separateAngular} shows the ``score'' of the hyperplane as a function of the number of features used. The score is the mean accuracy of training examples. In other words, it is the fraction of particles that fall on the proper side of the hyperplane, $S > 0$ for rearranging particles and $S < 0$ for non-rearranging particles. For many steps, the score is virtually unaffected as relatively inconsequential structure functions are pruned. At some point, the score begins to decrease and suddenly downturns once RFE reaches some crucial subset of features.

\begin{figure}
 	\includegraphics[width=\linewidth]{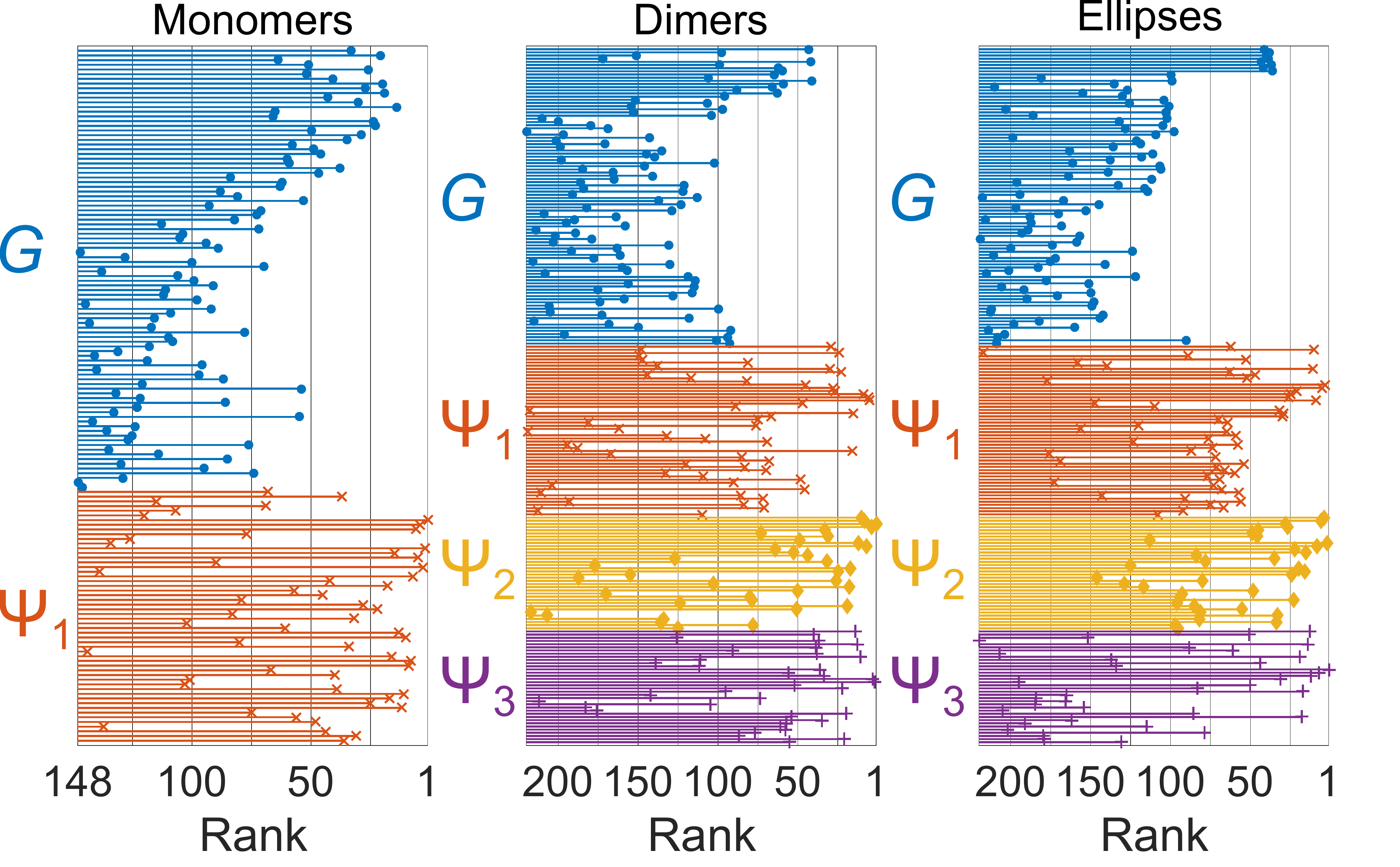}
	\caption{\label{fig:featureRanking_Angular} Stem plots of the structure functions as ranked by RFE for monomers, dimers and ellipses, grouped by structure function type. Vertical gridlines guide the eye to identify groups of structure functions that rank similarly.}
\end{figure}

One wrinkle of RFE not explored here is that a variable number of features could be removed at each iteration. In practical applications, this can aid in the reduction of a SVM with very high dimensionality. When we eliminate structure functions one at a time, this also yields a ranking of the structure functions by relative importance. The resulting rankings we find for the three shapes are shown in Fig.~\ref{fig:featureRanking_Angular}. Tables~\ref{table:MonomerRanking},~\ref{table:DimerRanking}, and~\ref{table:EllipseRanking} explicitly list the top 10 ranking features for monomers, dimers, and ellipses, respectively. It is vital to mention that there is no expectation for these rankings to be robust. That is to say, we are not interested in proving that the top-ranking feature is definitively more important than the second, which is definitively more important than the third, and so on. However, these rankings are valuable in highlighting groups of features that are collectively important in determining $S$. Furthermore, it offers a path along which to explore the effects of individual structure functions, without guessing from a large collection.

% Top ten structure functions (separate angular, orientational)
% Monomers: 101, 107, 111, 102, 109, 103, 113, 131, 132, 126
\begin{table}[ht]
\centering
\begin{ruledtabular}
\caption{Top ranked features for monomers.}
\begin{tabular}{ccc}
 Rank & Function & Neighbor Species \\ \hline
 1 & $\Psi_1\left(\xi = 2.554D, \zeta = 2, \lambda = -1\right)$ & LL \\
 2 & $\Psi_1\left(\xi = 1.648D, \zeta = 1, \lambda = +1\right)$ & LL \\
 3 & $\Psi_1\left(\xi = 1.648D, \zeta = 2, \lambda = -1\right)$ & LS \\
 4 & $\Psi_1\left(\xi = 2.554D, \zeta = 2, \lambda = -1\right)$ & LS \\
 5 & $\Psi_1\left(\xi = 1.648D, \zeta = 1, \lambda = +1\right)$ & SS \\
 6 & $\Psi_1\left(\xi = 2.554D, \zeta = 2, \lambda = -1\right)$ & SS \\
 7 & $\Psi_1\left(\xi = 1.204D, \zeta = 1, \lambda = +1\right)$ & LL \\
 8 & $\Psi_1\left(\xi = 0.933D, \zeta = 4, \lambda = +1\right)$ & LL \\
 9 & $\Psi_1\left(\xi = 0.933D, \zeta = 4, \lambda = +1\right)$ & LS \\
 10 & $\Psi_1\left(\xi = 0.933D, \zeta = 1, \lambda = +1\right)$ & LS \\
\end{tabular}
\label{table:MonomerRanking}
\end{ruledtabular}
\end{table}

\begin{figure}
 	\includegraphics[width=\linewidth]{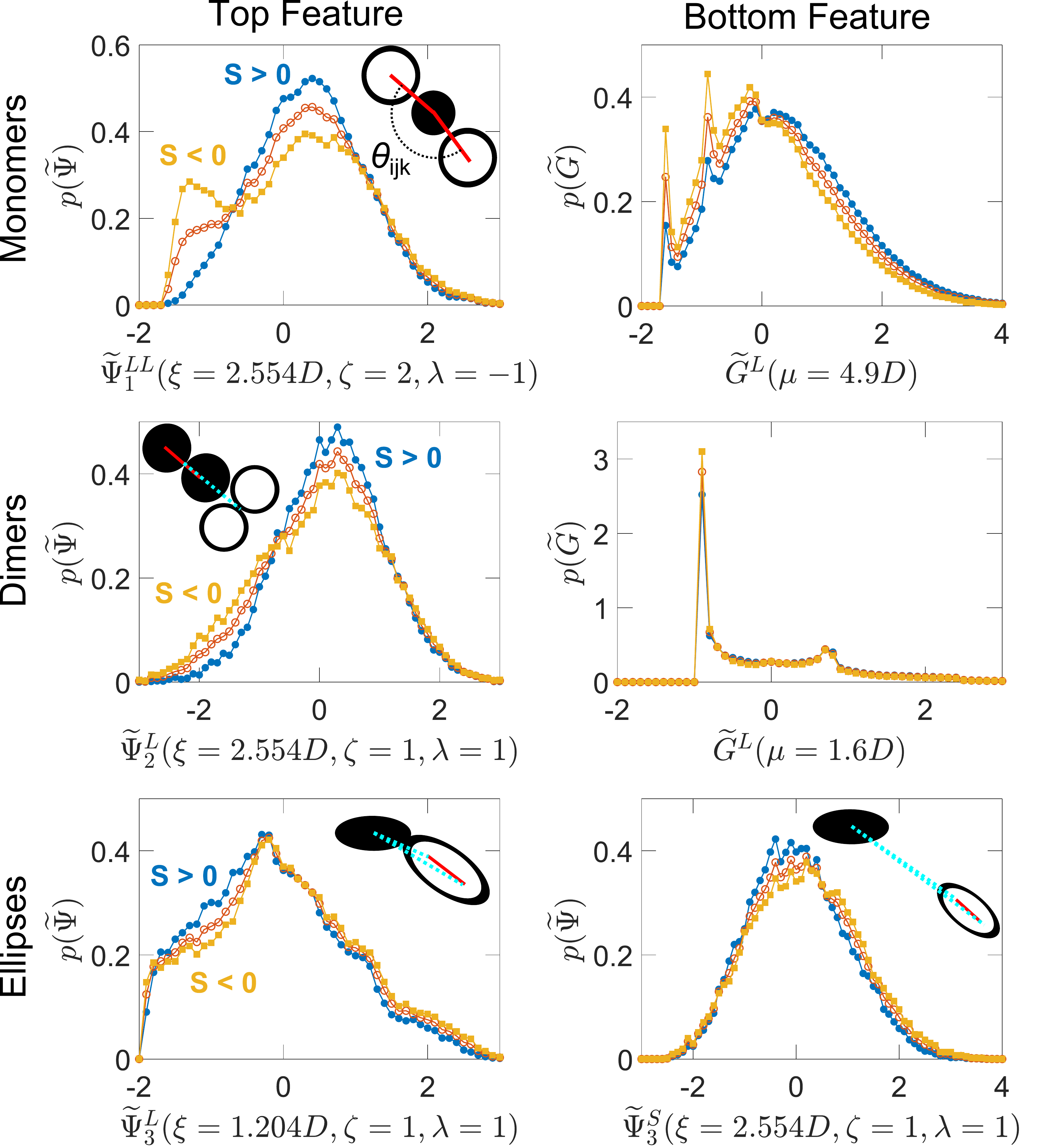}
	\caption{\label{fig:featureDistributions} Distributions of normalized values of the top- and bottom-ranked structure functions for monomers (top row), dimers (middle row), and ellipses (bottom row). The inset sketches present for distributions of angular and orientational structure functions illustrate the structural aspects that are quantified. The three curves in each plot correspond to all particles (open circles), those with $S > 0$ (closed circles), and those with $S < 0$ (closed squares). While the monomer distributions exhibit distinct behaviors in these three curves for the top-ranked feature, the others highlight the difficulty in using a singular structural measure to assess a particle's susceptibility to rearrangement.}
\end{figure}

The simplest way to consider effects of individual structure functions is to examine the top- and bottom-ranking structure function for each shape. Fig.~\ref{fig:featureDistributions} shows the distribution of structure function values for those at the top and bottom of the rankings. These plots also distinguish particles that are overall considered likely to rearrange, $S > 0$, and those considered less likely to rearrange, $S < 0$. Note that the structure function values are normalized, so that their values in the training set have zero mean and unit variance. 

For monomers in Fig.~\ref{fig:featureDistributions}(top row), there is a clear distinction in behaviors between the top- and bottom-ranking structure functions. Consider the top-ranking feature which, based on the parameter values for $\xi$, $\lambda$, and $\zeta$, emphasizes large angles that are swept between the reference monomer and nearby large monomers. The overall distribution for all monomers is roughly bimodal, with a clear peak near zero and a plateau at negative values. However, the distribution looks very different when only considering $S > 0$ or $S < 0$ monomers. The distributions of the bottom-ranking monomer structure function, meanwhile, are very similar whether we consider all monomers or split them by the sign of $S$. 

We cannot use the bottom-ranking structure function to interpret information about local rearrangements, but we can look at the top-ranking one. For $S < 0$, the peak at negative values is emphasized. Since negative values are more favored by comparison, this indicates that structural aspects described by that function are less prevalent. Meanwhile, this negative peak vanishes for $S > 0$ monomers. This indicates that many, but not all, particle rearrangements involve a monomer getting caught diametrically between large monomers. The subsequent rearrangement could involve the central monomer getting squeezed out, or perhaps one of the neighboring monomers sliding past. 

% Dimers: 151, 201, 200, 152, 112, 111, 158, 150, 110, 149
\begin{table}[ht]
\centering
\begin{ruledtabular}
\caption{Top ranked features for dimers, when softness is computed with separate angular and orientational features.}
\begin{tabular}{ccc}
 Rank & Function & Neighbor Species \\ \hline
 1 & $\Psi_2\left(\xi = 2.554D, \zeta = 1, \lambda = +1\right)$ & L \\
 2 & $\Psi_3\left(\xi = 1.204D, \zeta = 4, \lambda = +1\right)$ & L \\
 3 & $\Psi_3\left(\xi = 1.204D, \zeta = 2, \lambda = +1\right)$ & S \\
 4 & $\Psi_2\left(\xi = 2.554D, \zeta = 1, \lambda = +1\right)$ & S \\
 5 & $\Psi_1\left(\xi = 1.648D, \zeta = 2, \lambda = +1\right)$ & SS \\
 6 & $\Psi_1\left(\xi = 1.648D, \zeta = 2, \lambda = +1\right)$ & LS \\
 7 & $\Psi_2\left(\xi = 1.648D, \zeta = 1, \lambda = +1\right)$ & S \\
 8 & $\Psi_2\left(\xi = 2.554D, \zeta = 1, \lambda = -1\right)$ & S \\
 9 & $\Psi_1\left(\xi = 1.648D, \zeta = 2, \lambda = +1\right)$ & LL \\
 10 & $\Psi_2\left(\xi = 2.554D, \zeta = 1, \lambda = -1\right)$ & L \\
\end{tabular}
\label{table:DimerRanking}
\end{ruledtabular}
\end{table}

While informative physics can be drawn out of the distribution of the top-ranking monomer struction function, the same cannot be said of dimers and ellipses. For either the top-ranking or bottom-ranking structure functions, none of the distributions appear especially different. Fig.~\ref{fig:RFE_scores_separateAngular} suggests why this is the case. While the score for monomers is about 60\% when only 1 feature, the top-ranking structure function, is used, the score for dimers and ellipses falls right near 50\%. This means using only one structure function for these shapes is about as useful as flipping a coin. We see that the score for dimers does jump to around 60\% when two structure functions are used, so we instead consider bivariate distributions of those two structure functions.

\begin{figure}
 	\includegraphics[width=\linewidth]{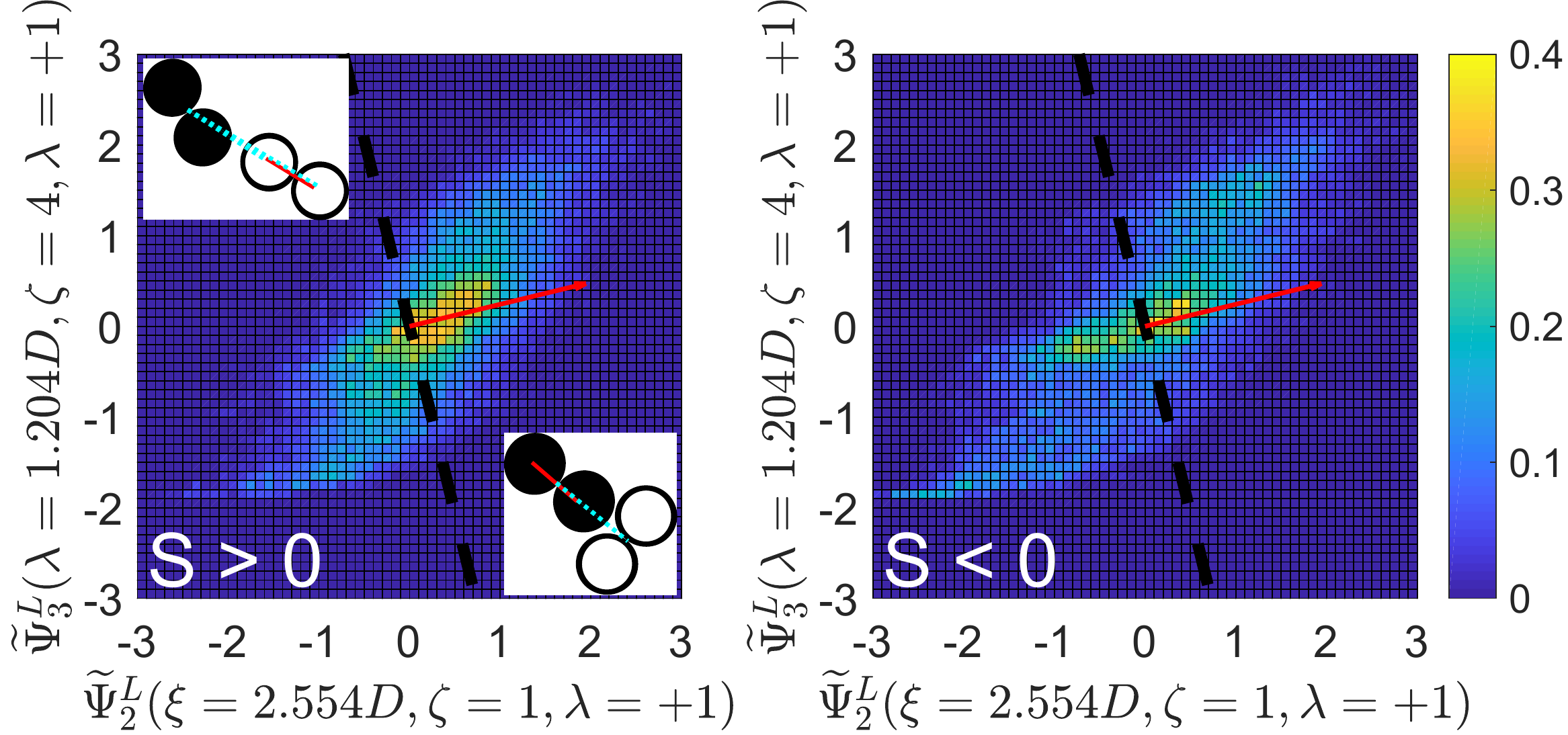}
	\caption{\label{fig:dimers_topTwo} Bivariate distributions of the top 2 structure functions for dimers, those with $S > 0$ on the left and those with $S < 0$ on the right. The inset sketches in the left plot illustrate the structural aspects that are quantified along each axis. The dotted line and arrow are projections of the full hyperplane and normal vector, respectively. While the top feature does not provide insight from Fig.~\ref{fig:featureDistributions}, the distribution of two features starts to illuminate structural aspects that are more likely to rearrange.}
\end{figure}

Fig.~\ref{fig:dimers_topTwo} shows the bivariate distributions for the top 2 ranking structure functions. On the left side, we show the distribution of normalized structure function values for dimers with $S > 0$, and those of dimers with $S < 0$ on the right side. In addition, we include the projection of the full hyperplane and its normal vector onto these axes. Visualizing the structure function values in this way now shows how the collective values of these two functions trends well with overall $S$ values. When $S > 0$, there is a dense cluster where both functions have positive values, where there is an abundance of these structural aspects relative to the mean.  Conversely, when $S < 0$, this dense cluster vanishes, and is replaced by a plateau that lies directly opposite the normal vector projection. The collective physical interpretation of these observations is to consider dimers whose position vectors to neighboring dimers form small angles, both with respect to the reference dimer orientation and the orientation of the neighboring dimer. In other words, dimers that form tip-to-tail formations seem to be especially likely to rearrange.

% Ellipses: 197, 157, 107, 149, 108, 150, 198, 158, 112, 96
\begin{table}[ht]
\centering
\begin{ruledtabular}
\caption{Top ranked features for ellipses, when softness is computed with separate angular and orientational structure functions.}
\begin{tabular}{ccc}
 Rank & Function & Neighbor Species \\ \hline
 1 & $\Psi_3\left(\xi = 1.204D, \zeta = 1, \lambda = +1\right)$ & L \\
 2 & $\Psi_2\left(\xi = 1.648D, \zeta = 1, \lambda = +1\right)$ & L \\
 3 & $\Psi_1\left(\xi = 1.648D, \zeta = 1, \lambda = +1\right)$ & LL \\
 4 & $\Psi_2\left(\xi = 2.554D, \zeta = 1, \lambda = -1\right)$ & L \\
 5 & $\Psi_1\left(\xi = 1.648D, \zeta = 1, \lambda = +1\right)$ & LS \\
 6 & $\Psi_2\left(\xi = 2.554D, \zeta = 1, \lambda = -1\right)$ & S \\
 7 & $\Psi_3\left(\xi = 1.204D, \zeta = 1, \lambda = +1\right)$ & S \\
 8 & $\Psi_2\left(\xi = 1.648D, \zeta = 1, \lambda = +1\right)$ & S \\
 9 & $\Psi_1\left(\xi = 1.648D, \zeta = 2, \lambda = +1\right)$ & SS \\
 10 & $\Psi_1\left(\xi = 2.554D, \zeta = 1, \lambda = -1\right)$ & LS \\
\end{tabular}
\label{table:EllipseRanking}
\end{ruledtabular}
\end{table}

While some physical intuition can be made for monomers and dimers with just one or a couple structure functions, the same cannot be said of ellipses, unfortunately. According to Fig.~\ref{fig:RFE_scores_separateAngular}, the score of ellipses does not substantially improve until several structure functions are included. While these structure functions are explicitly listed in Table~\ref{table:EllipseRanking}, it is impossible to visualize in this article their collective effects, while also providing simplified physical intuition. We see that both angular and orientational structure function types are included, some forming large angles, others small, and with varying species of neighboring particles. While this aspect of our study of ellipses does not reduce as simply as that of monomers and dimers, this observation underlines the original thesis of the SVM approach. In order to identify constituents of a disordered solid that are likely to rearrange, one can consider a model that weighs the contribution of several structural considerations at once. Relying on individual structural metrics can be presumptuous. However, that is not to say there are other untested structural measurements that can be applied to ellipses that could prove to be discriminatory.

\section{Discussion}
\label{sec:Discussion}

In this article, we presented a computational approach to characterizing structural defects in real disordered solids of elongated particles. Using granular pillars as our model solid, we studied systems made up of circularly symmetric monomers, concave dimers, and convex ellipses. We followed a SVM approach that has been used by other researchers to characterize defects in a wide range of disordered solids. The final output is a particle-scale parameter, softness, that serves as a continuous quantity correlating with the susceptibility of a particle to rearrangement.

The original SVM approach relies on structure functions defined by Behler and Parrinello to quantify how particle centers are dispersed radially and angularly. These functions do not account for orientations of elongated particles. However, following their original definitions, we have devised orientational structure functions to capture this missing aspect of structure. Once we include these new structure functions, dimers exhibit similar predictive performance as monomers, and ellipses perform even better. The discrepancy in performance between ellipses and both monomers and dimers remains an interesting question. Some of the difference could be a result of shape effects. The rounded convex shape of ellipses ensures there are no interlocking effects, unlike dimers. Monomers also have convex surfaces, and exhibit similar performance to ellipses using just the Behler-Parrinello structure functions. Perhaps in the absence of interlocking contacts, the radial and orientational structure functions provide a meaningful collection of structure, which can only be improved upon with the introduction of orientational structure functions. We also cannot discount the possibility of material effects, since the ellipses are fabricated from a light-cured polymer while the monomers and dimers are made of acetal.

Using the final softness fields computed for each shape, with separate angular and orientational structure functions, we explore the physics inherent to this field, and how it relates to the rearrangement field, $D^2_{\mathrm{min}}$. In particular, we examine spatial correlations in the $D^2_{\mathrm{min}}$, $S$, and $\Delta S$  fields and see that for every shape, autocorrelations and cross-correlations exhibit an exponential dependence with a characteristic length. This observation is directly in line with disordered solids in general, and now extends to anisotropic constituent particles. While both rearrangement and softness fields of monomers and dimers are highly localized, ellipses exhibit a substantially larger correlation length for softness compared to rearrangements. Again, this difference likely originates from a combination of shape and material effects. The elliptical shape allows for larger contact areas, which can encourage local orientational ordering, reducing structural variance in space. At the same time, changes in softness remain highly localized for all shapes. We believe the collective results for ellipses play a role in global ``barrelling'' behavior that is distinct from monomers and dimers. At the same time, physical properties of the ellipse acrylate material, e.g. friction, also differ from the acetal that make up monomers and dimers. Decoupling how these attributes affect pillar deformation is an open question that warrants further investigation. We are also intrigued by the observation of larger cross-correlation lengths compared to autocorrelation lengths for all shapes. Our current results indicate there may exist feedback behavior where a rearrangement causes a region to lose softness, while potentially creating another defect further away. This possibility warrants further study, particularly in simulations in which larger system sizes that suppress noise are more tractable. 

Finally, we demonstrated how another established machine learning tool, RFE, can be used in junction with our SVM approach to identify structural aspects that most---and least---influence the overall softness value. In the case of monomers and dimers, we successfully reduced structural considerations to one or two structural aspects, providing an avenue for one to begin comprehending these inherently high-dimensional calculations. An important aspect of material design is to consider the expected strain fields under mechanical and thermal loads. Is it possible to construct a disordered solid with a desired softness field, such that fractures or plastic deformations can be mitigated? In some contexts, would such aspects of design cross the threshold between brittle and ductile fracture behavior? RFE provides initial structural considerations to consider first, which may provide the basis for an algorithmic packing protocol to achieve these and similar goals.

Of course, there are other fundamental questions regarding the softness field that should be addressed, particularly surrounding time evolution. How does $S$ evolve coming into and out of individual rearrangements? Much of the work done across disordered solids focuses on yield behavior, how proportional changes in the softness field at yield is similar across various systems. Is there universal behavior in these systems beyond yield? Similar sizes in rearrangements and defects across these systems would indicate that there is. 

Even the construction of softness is still under consideration, specifically in terms of the choice of structure functions. The Behler-Parrinello structure functions were originally developed in the context of fitting potential energy surfaces to atomic positions~\cite{BehlerParrinelloPRL2007}. Their choice is not unique, and several other descriptors of atomic environments have recently been reviewed~\cite{BartokPRB2013}. Alternative structure functions can derive from bond-order parameters, power spectra, and Fourier series of structural invariants. Analogous versions of some these descriptors could also prove useful for amorphous packings of elongated particles.

As mentioned in Sec.~\ref{sec:Intro}, there is a wealth of literature regarding structural anisotropies in particulate systems, often measured with relation to the Voronoi tessellation. There also exist structural metrics that partitition the structure with methods other than Voronoi tessellation, some under current development. For example, relative angular position~\cite{VanMeelJCP2012,Das2018} serves as a way to define nearest neighbors without distance thresholds or Voronoi cells. Also, a major focus of recent research relates concepts of network science to grains in contact~\cite{NetworksGrainsReview2018}. In particular, a local metric can be extracted that describes the centrality of a particle within the overall structure~\cite{Kollmer2018}. It is entirely possible that these metrics may serve as viable alternatives to the Behler-Parrinello radial and angular structure functions, and/or our newly defined orientational structure functions. Moreover, an optimal subset of structural features could include a combination of structural measurements that come from multiple perspectives.
%Falk makes an interesting point in his 2018 PRE about how softness, and other structural signatures, do not incorporate the orientations of STZs; ours here is still a scalar, but perhaps the framework here can inform how one might develop a ``softness vector"

\section*{Acknowledgments}
We thank J. M. Rieser for technical assistance, F. P. Landes for analysis assistance, R. J. S. Ivancic for helpful discussions, and Walt Barger for fabrication assistance. This work was supported by NSF Grant MRSEC/DMR-1720530. 

\bibliography{ML_Paper_bib}

\end{document}